# Dielectric properties of graphene on transition metal dichalcogenide substrate

**Partha Goswami**

Physics Department, D.B.College, University of Delhi, Kalkaji, New Delhi-110019, India

**Abstract** The tunability aspect of the dielectric properties induced by the substrate driven interactions (SDI) and the exchange field (M) due to the ferro-magnetic impurities in graphene monolayer on transition metal dichalcogenide is reported here. The interactions involve sub-lattice-resolved, enhanced intrinsic spin-orbit couplings(SOC), the extrinsic Rashba spin-orbit coupling (RSOC), and the orbital gap related to the transfer of the electronic charge from graphene to the substrate. We obtain the gapped bands with a RSOC-dependent pseudo Zeeman field due to the interplay of SDI. This enables us to obtain an expression of the dielectric function in the finite doping case ignoring the spin-flip scattering events completely. We find that the stronger RSOC has foiling effect on the Thomas-Fermi screening length.This foiling effect, over a broad range of the exchange field values (0– 1meV), is an indication of the domination of the Dyakonov-Perel mechanism in the system over the Elliot-Yafet spin relaxation mechanism.The zero of the dielectric function corresponds to the plasmon frequency. We find that there is only one such frequency. The Plasmon dispersion yields the $q^{2/3}$ behavior and not the well known $\sqrt{q}$ behavior. We also find that the plasmon frequency could be changed by the tuning of the chemical potential.

**I. INRODUCTION** The isolation and production of the graphene in 2004[1,2] –a two-dimensional material with a single layer honeycomb lattice of carbon atoms– have generated substantial theoretical and experimental research activity[2]. As a material for application in the future nano-scale electronics and photonics, it has attracted great attention world-wide. The graphene was found to possess host of unusual properties, such as the high carrier mobility, the low resistivity, the large in-plane stiffness[3], the larger than metal optical absorption at the Plasmon-resonance [4], and so on. The tunable plasmonic materials, like graphene, are required in various areas of photonics. Their applications range from broadband optical modulators, micro-ring resonator, nano-resonators, tunable terahertz meta-materials, tunable terahertz hybrid metal-graphene structure, real-time tunable lasing from nanocavity arrays, perfect absorber materials, and radiators to highly practicable (bio) chemical sensing [5-16]. The optical absorption (nearly 2.3% from the visible to near-infrared spectral windows) of pristine, pure graphene monolayer is featureless as the optical conductivity corresponds to a constant value $\sigma_0 = e^2/4\hbar$ which is independent of any material parameters [17]. Such feature-less-ness is a manifestation of the fact that Dirac fermions here are mass-less. However, these optical properties of graphene can be remarkably altered by changing the Fermi energy with the further addition of charge carriers [18].For example, doped graphene supports plasmons that are tunable, providing a novel platform for tunable devices. The large mobility of charge carriers in graphene makes high-speed tunable plasmonics possible by electrical gating of graphene.

Despite these remarkable properties, it has not been possible to fully exploit the graphene's potential due to the difficulty of opening a reasonably sized insulating gap in its band structure. The absence of the gap is owing to graphene's weak spin-orbit coupling.It was shown by Kane and Mele [19] that the gap opens up in the spectrum if one includes intrinsic spin-orbit interactions(SOI)[19]. The spin degeneracy of the spectrum could be lifted through the Rashba SOI which depends upon an external electric field. The experimental finding of a gap of 0.26 eV when graphene is epitaxially grown on the SiC substrate[20] shows the way to yet another mode of gap ($\Delta_{orbital}$) opening. This gap increases as the sample thickness decreases. It has been proposed that the origin of this gap is the breaking of sub-lattice symmetry owing to the graphene-substrate interaction. In this communication we study the interesting and useful possibilities related to the dielectric properties of graphene on TMDC substrate. The possibilities follow on the heels of the engineering of the enhanced spin-orbit coupling (SOC) in graphene through interfacial effects via coupling to the suitable substrates, viz. a two dimensional transition metal dichalcogenide (TMDC). The graphene layer is also exchange(*M*)coupled to the magnetic impurities, e.g. Fe atoms deposited to the graphene surface. A direct, functional electric field control of magnetism at the nano-scale is needed for the effective demonstration of our results related to the exchange-field dependence. The magnetic multi-ferroics, like $BiFeO_3$ (BFO) have piqued the interest of the researchers world-wide with the promise of the coupling between the magnetic and electric order parameters.

**TABLE 1.** The values of the substrate-induced interactions for graphene on the TMDC WSe$_2$, WS$_2$, MoSe$_2$, and MoS$_2$.

| TMDC | $t$ [eV] | $\Delta_{Orbital}$ [meV] | $t_{so}^A$ [meV] | $t_{so}^B$ [meV] | $\lambda'_R$ [meV] |
|---|---|---|---|---|---|
| WSe$_2$ | 2.51 | 0.54 | −1.22 | 1.16 | 0.56 |
| WS$_2$ | 2.66 | 1.31 | −1.02 | 1.21 | 0.36 |
| MoSe$_2$ | 2.53 | 0.44 | −0.19 | 0.16 | 0.26 |
| MoS$_2$ | 2.67 | 0.52 | −0.23 | 0.28 | 0.13 |

A large number of theoretical investigations of the dielectric function of graphene-based systems have been reported in the past few years. These include gapped graphene**[21-23],** multilayer graphene samples**[24-29]**, graphene under a circularly polarized electro-magnetic field**[30]**, and the graphene antidot lattices **[31].** In this paper, as already mentioned, we examine the dielectric properties of graphene including four substrate driven interactions and the exchange field. The problem is novel one as the behavior of the dielectric function in the presence of such a large number of interactions is unknown. The four substrate-induced interaction terms correspond to (i) the orbital gap related to the transfer of the electronic charge from graphene to TMDC,(ii) the sub-lattice-resolved, giant intrinsic spin-orbit couplings(SOC) due to the hybridization of the carbon orbitals with the d- orbitals of W/ Mo, and (iii) the extrinsic Rashba spin-orbit co-upling that allows for external tuning of the band gap in graphene and connects the nearest neighbours with spin-flip. These interactions are time-reversal invariant and absent by inversion symmetry in isolated pristine, pure graphene monolayer. The report **[32]** on CVD graphene samples and graphene in proximity to WS$_2$ have shown appreciable spin-orbit coupling which are induced either by defects or by the proximity to materials with high spin-orbit coupling. The evolution of the band structure of graphene introducing these interactions have been discussed by Gmitra et al.**[33-36]** extensively. The Zeeman field, however, was conspicuous by its absence in their analysis. This is very much required here as the field provides a platform to us to justify the domination of the Dyakonov-Perel mechanism **[37]** over the Elliot-Yafet mechanism **[38]** in the calculation of the Thomas-Fermi screening length and the plasmon frequency in section 3. Some of the values of the orbital and spin-orbital parameters are summarized in table 1 as well. These parameters can be tuned by a transverse electric field and vertical strain. As could be seen in this table, the sum of the absolute value of the intrinsic SOC terms is greater than the term $\Delta_{Orbital}$ characterizing the (staggered) sub-lattice asymmetry in the graphene A and B atoms on WSe$_2$ and WS$_2$ whereas it is less for MoY$_2$. It could be seen **[33-36]** that as long as the former is valid the anti-crossing of bands with opposite spins takes place, due to the presence of the Rashba term, around each of the valleys near the **K** point of graphene, However, when the latter is true, one makes a cross-over to a 'direct band' regime with typically parabolic dispersion for each of the two spin projections.

Upon going back to the frame of this article, we mention that the Zeeman-like term of the spectrum in Eq.(10) below appears due to the interplay of the substrate induced interactions with the prime player as the Rashba SOC. We re-iterate that the preliminary investigation of the interplay of these perturbations and the ferro-magnetic exchange field in the collective mode (dielectric properties) of the system is our main task. The detailed discussion on the consequences of this Zeeman field could be taken up elsewhere. We, however, emphasize that this field encourages the spin precession due to the effective magnetic field in the system over the spin-flip scattering of electrons due to momentum scattering. Thus, the field enables us to obtain an expression of the dielectric function in the finite doping case ignoring the spin-flip scattering events completely. The SOC interactions and the exchange field are in the band and do not act as scatterers here. It may be mentioned that inclusion of the exchange field effect in the band is not unprecedented. Introducing in the same manner, Macdonald et al.**[36]** have discussed the evolution of the electronic structure as the exchange field and Rashba SO coupling are introduced to the system. Our broad aim behind the investigation of the effect of the exchange field (*M*) is that, using graphene as a prototypical 2D system, we wish to see how material (and, in particular, dielectric) properties change under the influence of *M*. This is expected to pave the way to efficient control of spin generation and spin modulation in 2D devices without compromising the delicate material structures.

The paper is organized as follows: In Sec. II, we introduce the general form of low-energy graphene(monolayer on two dimensional transition metal dichalcogenides) Hamiltonian accounting for interaction with the substrate. The single-particle excitation spectrum spectrum is obtained from a quartic involving all the substrate induced perturbations.The Sec. III contains an investigation report of the dielectric properties of the system. The paper ends with discussion and concluding remarks in section IV.

## II. Hamiltonian with substrate induced interactions

The Hamiltonian (H) of the graphene monolayer on two dimensional transition metal dichalcogenides (such as, $WSe_2/WS_2/MoSe_2/MoS_2$) substrate is the starting point in this section. The Hamiltonian (H) is built on the orbital Hamiltonian ($H_0$) for pristine graphene. The former, apart from $H_0$, comprises of the staggered potential term ($H_\Delta$) describing the effective orbital energy difference on A and B sub-lattices of graphene, the sub-lattice-resolved intrinsic spin-orbit coupling($H_{SO}$)–a next-nearest neighbour hop-ping much larger than that, say, in the hydrogenated graphene,the pseudo-spin inversion asymmetry related spin-orbit coupling term– a next-nearest neighbor hopping with the spin-flip, and the Rashba-type spin-orbit coupling ($\lambda'_R$) which is actually the nearest-neighbor spin-flip hopping. The last term takes care of the spin-splitting away from the Dirac points **K** and **K′**. Furthermore, as already mentioned, ferromagnetic(FM) impurity atoms are deposited to the graphene surface. Since FM results from the interaction of the atomic moments in materials, there is an exchange energy associated with coupling the spin moments. The exchange interaction is usually replaced by a spin dipole moment and the Weiss field : $H_{ex} = M\, s_z$, where the spin index $s_z = \pm 1$. In the case of iron, it is nearly 1.1 meV. We mention that the FM impurities do not act as scatterer in our scheme; their effect is included in the band dispersion. This modus operandi to extract the exchange coupling effect has been used in the case of graphene and silicene by several authors **[36,40-42]**. There is also an indirect coupling, often referred to as the Ruderman-Kittel-Kasuya-Yosida (RKKY) interaction**[39]**, which couples moments over large distance. It acts through an interm-ediary which in metal are itinerant electrons. Since it is a dominant interaction only in metal, we shall not be considering here. The term H is basically a low-energy effective Hamiltonian around **K** and **K′** in the basis (A↑, B↓, A↓, B↑) or ($a^\xi_{k\uparrow}$, $b^\xi_{k\downarrow}$, $a^\xi_{k\downarrow}$, $b^\xi_{k\uparrow}$) in momentum space. Here $a^\xi_{k\sigma}(b^\xi_{k\sigma})$ is the fermion annihilation operator for the state ($k,\sigma$) corresponding to the valley $\xi = \pm 1$, and the sub-lattice A(B). Thus, the low-energy Hamiltonian **[33-36]** for the graphene on TMDC system may be written down explicitly as

$$H/\left(\frac{\hbar v_F}{a}\right) = \sum_{\delta k} (a^{\dagger\xi}_{\delta k\uparrow}\ b^{\dagger\xi}_{\delta k\downarrow}\ a^{\dagger\xi}_{\delta k\downarrow}\ b^{\dagger\xi}_{\delta k\uparrow}) \frac{\hbar((\delta k))}{\left(\frac{\hbar v_F}{a}\right)} \begin{pmatrix} a^\xi_{\delta k\uparrow} \\ b^\xi_{\delta k\downarrow} \\ a^\xi_{\delta k\downarrow} \\ b^\xi_{\delta k\uparrow} \end{pmatrix} \qquad (1)$$

$$\frac{\hbar(\delta k)}{\left(\frac{\hbar v_F}{a}\right)} = \begin{pmatrix} a_1 & h^\xi_+ & -(\lambda^+ + \lambda^-)ia\delta k_- & -a\delta k^\xi_+ \\ h^{\xi*}_+ & a_2 & -a\delta k^\xi_- & -(\lambda^+ - \lambda^-)ia\delta k_+ \\ (\lambda^+ + \lambda^-)i\,a\delta k_+ & -a\delta k^\xi_+ & a_3 & h^\xi_- \\ -a\delta k^\xi_- & (\lambda^+ - \lambda^-)i\,a\delta k_- & h^{\xi*}_- & a_4 \end{pmatrix}. \qquad (2)$$

$$h^\xi_+ = 1.5i\lambda_R(E)(1+\xi),\ h^\xi_- = 1.5i\,\lambda_R(E)(1-\xi),$$

$$a_1 = \Delta + M + \xi\,\Delta^A_{soc},\ a_2 = -\Delta - M + \xi\,\Delta^B_{soc},\ a_3 = \Delta - M - \xi\,\Delta^A_{soc},$$

$$a_4 = -\Delta + M - \xi\,\Delta^B_{soc} \qquad (3)$$

Here the nearest neighbor hopping is parameterized by a hybridization $t, \left(\frac{\hbar v_F}{a}\right) = \left(\frac{\sqrt{3}}{2}t\right)$, and $a = 2.46$ A° is the pristine graphene lattice constant. The various terms in the Hamiltonian is made dimensionless dividing by the energy term $\left(\frac{\hbar v_F}{a}\right)$. Also, the dimensionles momenta $a\delta k^\xi_\pm \rightarrow a\delta k_\pm$ (that is, $a\delta k_x \pm i\, a\delta k_y$) for the Dirac point **K** ($\xi = +1$) and $a\delta k^\xi_\pm \rightarrow a\delta k^*_\pm$ (that is, $a\delta k_x \mp i\, a\delta k_y$) for the Dirac point **K′** ($\xi = -1$). The fields ($a_1, a_2, a_3, a_4$) could also be written down as $E(s_z, t_z) = \xi\, t'_{so}\, s_z t_z + \Delta\, t_z + M\, s_z$, with the spin index $s_z = \pm 1$ and the sub-lattice pseudo-spin index $t_z = \pm 1$. The parameters orbital proximity gap $\Delta$, the intrinsic parameters $\Delta^A_{soc}$ and $\Delta^B_{soc}$, and the extrinsic Rashba parameter $\lambda_R(E_z)$ allow for tuning by the applied electric field. Since the TMDC layer provides different environment to atoms A and B in the graphene-cell, there is (dimension-less) orbital proximity gap $\Delta = \Delta_{Orbital}/\left(\frac{\hbar v_F}{a}\right)$ arising from the effective staggered potential induced by the pseudo-spin symmetry breaking. The orbital gap $\Delta_{Orbital}$ is about 0.5 meV **[33-35]** in the absence of electric field. When the field crosses a limiting value 0.5 V/nm, the gap exhibits a sharp increase. This gap is related to the transfer of the electronic charge from graphene to TMDC. Due to the hybridization of the carbon orbitals with the d-orbitals of W/Mo, there is sub-lattice-resolved, giant intrinsic spin-orbit couplings($t_{so}^A$, $t_{so}^B$) : $\Delta^A{}_{soc}= \frac{t_{so}^A}{\left(\frac{\hbar v_F}{a}\right)}$ , $\Delta^B{}_{soc}= \frac{t_{so}^B}{\left(\frac{\hbar v_F}{a}\right)}$. These couplings correspond to next-nearest neighbor hopping[33-35]without spin-flip. The spin-orbit field parameters for graphene on TMDC are about 0.60 meV, which is over 20 times greater than that in pure graphene [33-36] ($t_{soc}$ ~ 24 μeV). The parameter $\lambda_R = \lambda'_R$ / ($\hbar v_F/a$) is the extrinsic Rashba spin-orbit coupling (RSOC), that allows for external tuning of the band gap in the system and connects the nearest neighbors with spin-flip. It, thus, arises because the inversion symmetry is broken when graphene is placed on top of a TMDC. While the intrinsic parameters $\Delta^A{}_{soc}$ and $\Delta^B{}_{soc}$ change rather moderately with the increase in the applied electric field, the Rashba parameter $\lambda_R$ almost doubles in increasing the field from −2 to 2 V/nm. For the pristine graphene $\lambda'_R \approx$ 10 μeV whereas for GTMDC(WSe$_2$) it is 0.56 meV. Wang et al.[36], however, have reported it to be approximately 1 meV. The sub-lattice resolved, pseudo-spin inversion asymmetry(PIA)driven spin-orbit coupling term, on the other hand, represents the next-nearest-neighbor, unlike the Rashba term, same sub-lattice hopping away from **K** and **K′** albeit with a spin flip. In the basis ($a_{\mathbf{k}\uparrow},b_{\mathbf{k}\uparrow},a_{\mathbf{k}\downarrow},b_{\mathbf{k}\downarrow}$),the ASOC terms, involving $\lambda^+ = \lambda'^+/(\frac{\hbar v_F}{a})$ , $\lambda^- = \lambda'^- / \left(\frac{\hbar v_F}{a}\right)$, could be written in a manner as shown in Eq.(2). Here $\lambda'^+$ and $\lambda'^-$, respectively, are the spin-orbit interactions representing the average coupling, and the differential coupling between the A and B sub-lattices. The spin-splitting by the Rashba term away from the points **K** and **K′** is the same as that at **K** and **K′**. We have dropped $\lambda'^+$ and $\lambda'^-$ involving terms as these are found to be in the nature of corrections to the momentum dependent terms in the band structure. The three spin-orbit interaction terms, with coupling constant (**$t_{so}^A$, $t_{so}^B$**) and **$\lambda'_R$**, are induced by interfacial interactions. The all four substrate-induced interaction terms, $\Delta_{Orbital}$, (**$t_{so}^A$, $t_{so}^B$**) and **$\lambda'_R$**, are time-reversal invariant and absent by inversion symmetry in isolated graphene sheets. Almost all the above parameters can be tuned by a transverse electric field and vertical strain. We have not considered the intrinsic RSOC for the following reason: Unlike conventional semiconducting 2D electron gases, in which the Rashba coupling is modeled as $\alpha(\delta k_y \sigma_x − \delta k_x \sigma_y)$ where $\sigma's$ are the Pauli matrices, the Rashba coupling in graphene does not depend on the momentum. The reason is that Rashba coupling is proportional to velocity, which is constant for mass-less Dirac electrons in graphene.

The energy eigen-values ($E(a|\delta\mathbf{k}|)$) of the matrix (2) are given by a quartic considering all the four substrate-induced interaction terms. In terms of the powers of ε ( where ε ≡ $E(a|\delta\mathbf{k}|)/\lambda_R$)**,** in the absence of PIA driven terms, the quartic may be written as $\varepsilon^4 − 2\varepsilon^2 b − 4\varepsilon c + d = 0$ or, $\varepsilon^4 − 2\varepsilon^2 b + b^2 = 4\varepsilon c + b^2 − d$, where

$$b_\xi(|\delta\mathbf{k}|,M) = [\Delta^2 + \frac{(\Delta^A{}_{soc})^2 + (\Delta^B{}_{soc})^2}{2} + M^2 + (a|\delta\mathbf{k}|)^2 + (9/4)(1+\xi^2)\lambda_R{}^2 − \xi\{|\Delta^A{}_{soc}|+\Delta^B{}_{soc}\} M],$$

$$c_\xi(M) = −\xi\{|\Delta^A{}_{soc}|−\Delta^B{}_{soc}\}[(9/4)\lambda_R{}^2 − (\frac{\Delta}{2})\xi\{|\Delta^A{}_{soc}|+\Delta^B{}_{soc}\} + \Delta M], \quad d_\xi(|\delta\mathbf{k}|,M) = \sum_{j=1}^{4} d_j + (a|\delta\mathbf{k}|)^4,$$

$$d_1 = \{\Delta^2 − (\xi\Delta^B{}_{soc} − M)^2\} \times \{\Delta^2 − (\xi|\Delta^A{}_{soc}| − M)^2\},$$

$$d_2 = (\frac{9}{4})\lambda_R{}^2 (1+\xi)^2 [(\Delta − M)^2 + \xi\{|\Delta^A{}_{soc}|+\Delta^B{}_{soc}\}(\Delta − M) + |\Delta^A{}_{soc}||\Delta^B{}_{soc}|],$$

$$d_3 = (\frac{9}{4})\lambda_R{}^2 (1−\xi)^2 [(\Delta + M)^2 + \xi\{|\Delta^A{}_{soc}|+\Delta^B{}_{soc}\}(\Delta + M) + |\Delta^A{}_{soc}||\Delta^B{}_{soc}|],$$

$$d_4 = 2(a|\delta\mathbf{k}|)^2 [(\Delta^2 − M^2) + \xi\{|\Delta^A{}_{soc}|+\Delta^B{}_{soc}\} M − |\Delta^A{}_{soc}||\Delta^B{}_{soc}|]. \quad (4)$$

Upon diagonalization of the Hamiltonian comprising of the four above-mentioned interactions and the exchange coup-ling *M* together with the kinetic energy term, we have thus obtained a complicated expression for the energy dispersion for both the valleys from a quartic. An approximate energy dispersion is derivable from a bi-quadratic if we assume the magnitudes of ($t^A{}_{SO}$, $t^B{}_{SO}$) to be equal. This, however, will not be appropriate for the discussion on collective oscillations for the non-appearance of the Zeeman-like term.The eigenvectors corresponding to the eigenvalues of the matrix in (2), are given by

$$\Psi^{\xi}_{s,\sigma}(\delta k, M) = \begin{pmatrix} \psi^{\xi}_{1,s,\sigma}(\delta k, M) \\ \psi^{\xi}_{2,s,\sigma}(\delta k, M) \\ \psi^{\xi}_{3,s,\sigma}(\delta k, M) \\ \psi^{\xi}_{4,s,\sigma}(\delta k, M) \end{pmatrix}. \qquad (5)$$

where

$$(a|\delta k|)^2 \psi^{\xi}_{2,s,\sigma}(\delta k, M) = -\frac{[h^{\xi*}_{+}(a_3-\varepsilon)(a|\delta k|)^2 + h^{\xi}_{-}(a_1-\varepsilon)(a\delta k^{\xi}_{-})^2]}{[(a_2-\varepsilon)(a_3-\varepsilon)-(a|\delta k|)^2]} \psi^{\xi}_{1,s,\sigma}(\delta k, M),$$

$$(a|\delta k|)^2 \psi^{\xi}_{3,s,\sigma}(\delta k, M) = -\frac{[h^{\xi*}_{+}(a\delta k^{\xi}_{+})(a|\delta k|)^2 + h^{\xi}_{-}(a_1-\varepsilon)(a_2-\varepsilon)(a\delta k^{\xi}_{-})]}{[(a_2-\varepsilon)(a_3-\varepsilon)-(a|\delta k|)^2]} \psi^{\xi}_{1,s,\sigma}(\delta k, M),$$

$$\psi^{\xi}_{4,s,\sigma}(\delta k, M) = [-\frac{h^{\xi}_{+}}{(a\delta k^{\xi}_{+})^2} \frac{\{h^{\xi*}_{+}(a_3-\varepsilon)\}(a\delta k^{\xi}_{+}) + \{(a_1-\varepsilon)h^{\xi}_{-}\}(a\delta k^{\xi}_{-})}{[(a_2-\varepsilon)(a_3-\varepsilon)-(a|\delta k|)^2]}$$

$$+\frac{(a_1-\varepsilon)}{(a\delta k^{\xi}_{+})}] \psi^{\xi}_{1,s,\sigma}(\delta k, M). \qquad (6)$$

We now go back to the eigenvalue equation and add and subtract an as yet unknown variable z within the squared subtract an as yet unknown variable $z$ within the squared term $(\varepsilon^4 - 2\varepsilon^2 b + b^2)$:

$$(\varepsilon^2 - b + z - z)^2 = 4\varepsilon c + b^2 - d \qquad (7)$$

$$(\varepsilon^2 - b + z)^2 = 2z\varepsilon^2 + 4\varepsilon c + (z^2 - 2bz + b^2 - d). \qquad (7a)$$

The necessity of retaining the relatively small term ($4\varepsilon c$) in Eq.(7) will be clear towards the end. Upon retaining the term ($4\varepsilon c$), Eqs.(7) or (7a) evidently becomes a quartic in $\varepsilon$ whereas ignoring the term ($4\varepsilon c$) we shall get a bi-quadratic with values of '$\varepsilon$' given by $\varepsilon^2 \approx b \pm \sqrt{(b^2-d)}$. We shall see below that without the term ($4\varepsilon c$) a complete discussion of the Thomas-Fermi sceening and plasmonics in graphene on TMD, which is one of our tasks here, does not seem to be possible. The left-hand side of Eq.(7) or (7a) is a perfect square in the variable $\varepsilon$. This motivates us to rewrite the right hand side in that form as well. Therefore we require that the discriminant of the quadratic in the variable $\varepsilon$ to be zero. This yields $16c^2 - 8z(z^2 - 2bz + b^2 - d) = 0$

or, $$z^3 - 2bz^2 + (b^2-d)z - 2c^2 = 0. \qquad (8)$$

The cubic equation above has the discriminant function

$$D = 8b^3c^2 - 72bdc^2 + 4d(b^2-d)^2 - 108c^4. \qquad (9)$$

Since $D$ is positive as could seen from the Figures 1(a) and 2(a) above (we have plotted here $D$ as a function of ($a\delta k$) for $M=0$ for graphene on WSe$_2$), we definitely have real roots of Eq.(8). These roots, as functions of '$M$' (in meV), are shown below in Figures 1(b) and 2(b) for the **K** and **K'** points respectively. The roots corresponding to the uppermost line in Figure 1(b) and 2(b) are the appropriate ones as they are found to be real, rational, and, importantly, being of positive sign yields real eigenvalues. Suppose we denote this root by $z_0(a\delta k, M, \xi)$. We find that $z_0(a\delta k, M, \xi) = z_0(-a\delta k, M, \xi)$ (see Figures 1(c) and 2(c)). Using (7) and (8) one may then write $\varepsilon^2 = b - z_0 \pm \{\sqrt{(2z_0)}\varepsilon + \sqrt{(2/z_0)c}\}$, or, $\varepsilon^2 - \sqrt{(2z_0)}\varepsilon + (-b + z_0 - c\sqrt{(2/z_0)}) = 0$ and $\varepsilon^2 + \sqrt{(2z_0)}\varepsilon + (-b + z_0 + c\sqrt{(2/z_0)}) = 0$. These two equations basically yield the band structure

$$\varepsilon_{\xi,s,\sigma}(a|\delta k|, M) = [s\sqrt{(z_0/2)}\lambda_R + \sigma\{(a|\delta k|)^2 + \lambda_{-s}(\xi, M)^2\}^{1/2}],$$

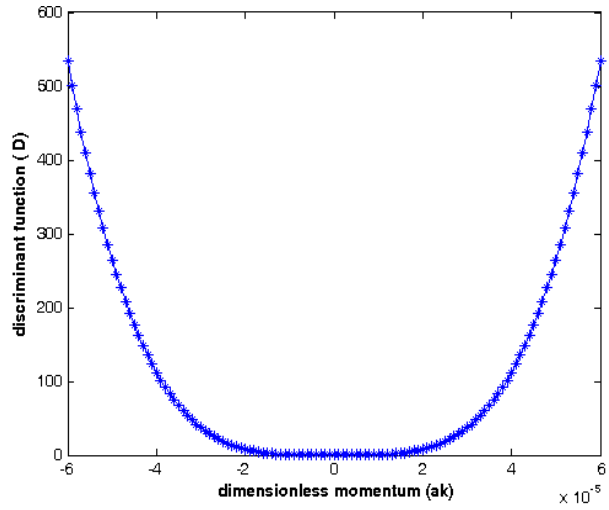

(a)

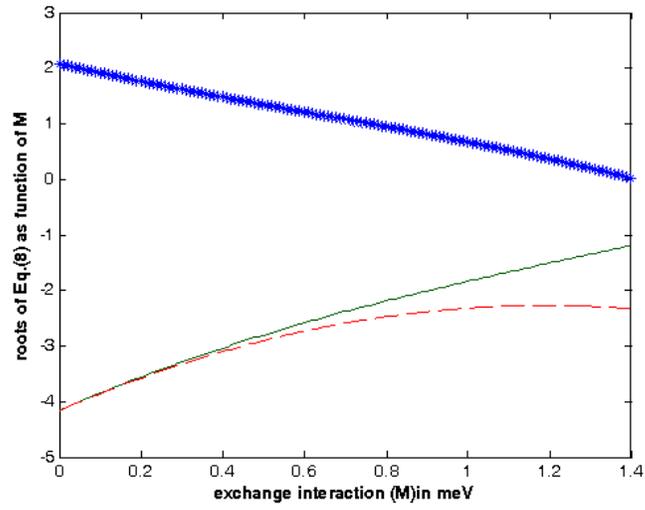

(b)

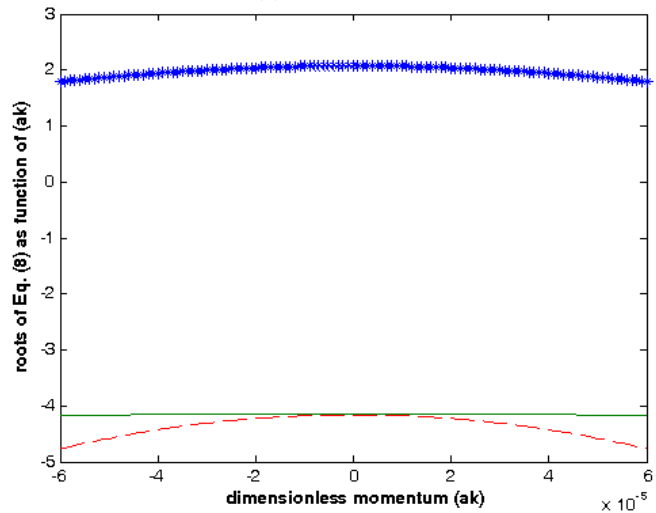

(c)

**Figure 1. (a)** A plot of the descriminant $\aleph$ as a function of $(a\delta k)$ at $M = 0$ for the Dirac point **K**. **(b)** The plots of the three (real and distinct) roots of (8) as functions of $M$ (in meV) at $(a\delta k) = 0$ for the point **K**. The uppermost curve corresponds to the admissible root $z_0(a\delta k, M, \xi)$ as this is positive. **(c)** The plots of the three (real and distinct) roots of Eq.(8) as functions of $(a\delta k)$ for $M = 0$ (for the point **K**). The blue line corresponds to $z_0(a\delta k, M, \xi)$. We find that $z_0(a\delta k, M, \xi) = z_0(-a\delta k, M, \xi)$. The plots correspond to graphene on $WSe_2$.

$$b_\xi(|\delta k|, M) = [\Delta^2 + (1/2)\{|\Delta^A_{soc}|^2 + \Delta^B_{soc}{}^2\} + M^2 + (a|\delta k|)^2 + (9/4)(1+\xi^2)\lambda_R^2 - \xi\{|\Delta^A_{soc}| + \Delta^B_{soc}\}M],$$

$$c_\xi(M) = -\xi\{|\Delta^A_{soc}| - \Delta^B_{soc}\}[(9/4)\lambda_R^2 - (\Delta/2)\xi\{|\Delta^A_{soc}| + \Delta^B_{soc}\} + \Delta M],$$

$$\lambda_s(\xi, M)^2 = \{\beta^2_\xi(M) - (z_0/2) + s\sqrt{(2c^2_\xi(M)/z_0)}\},$$

$$\beta^2_\xi(M) = [\Delta^2 + (1/2)\{|\Delta^A_{soc}|^2 + \Delta^B_{soc}{}^2\} + M^2 + (9/4)(1+\xi^2)\lambda_R^2 - \xi\{|\Delta^A_{soc}| + \Delta^B_{soc}\}M] \quad (10)$$

which consists of two spin-chiral conduction bands and two spin-chiral valence bands. Because of the spin-mixing driven by the Rashba coupling, the spin is no longer a good quantum number. Therefore, the resulting angular momentum eigenstates may be denoted by the spin-chirality index $s = \pm 1$. Here $\sigma = + (-)$ indicates the conduction (valence) band. The bands $\varepsilon_{\xi s, \sigma}(a|\delta k|, M)$ appear as the spin-valley resolved, and particle-hole symmetric. The latter follows from the fact that $z_0(a\delta k, M, \xi) = z_0(-a\delta k, M, \xi)$. Involving the spin-chiral index $s$, there is a Zeeman field like term $\approx (s\lambda_R\sqrt{(z_0(M,\xi)/2)}$ in Eq.(10)) due to the interplay of the substrate induced interactions with RSOC as the prime player. Thus, the (pseudo)spin of the electrons could be aligned by an RSOC induced pseudo magnetic field through this Zeeman term. It may be mentioned that the pseudo-spin alignment is also possible in graphene through a pseudo Zeeman field induced by mechanical deformation related vector potential. The field couples with different signs to states in the two valleys. Unlike this, our effective Zeeman field couples with same sign but different manner to the states in the valleys, as could be inferred from Figures 1 and 2. The appearance of the effective Zeeman field is a consequence of retaining the term $(4\varepsilon c)$ in Eqs.(7) or (7a). We have written in Eq.(10) above $b_\xi(a|\delta k|, M) = \varepsilon^2_{\delta k} + \beta^2_\xi(M)$ and the gap function $\lambda_s(\xi, M) = \{\beta^2_\xi(M) - (z_0/2) + s\sqrt{(2c^2_\xi(M)/z_0)}\}^{1/2}$. This gives $\varepsilon_{\xi s, \sigma}(a|\delta k| = 0, M) - (s\sqrt{(z_0/2)}\lambda_R) = \sigma\lambda_s(\xi, M)$. Thus, the function $\sigma\lambda_s(\xi, M)$ is actually the valence and conduction band

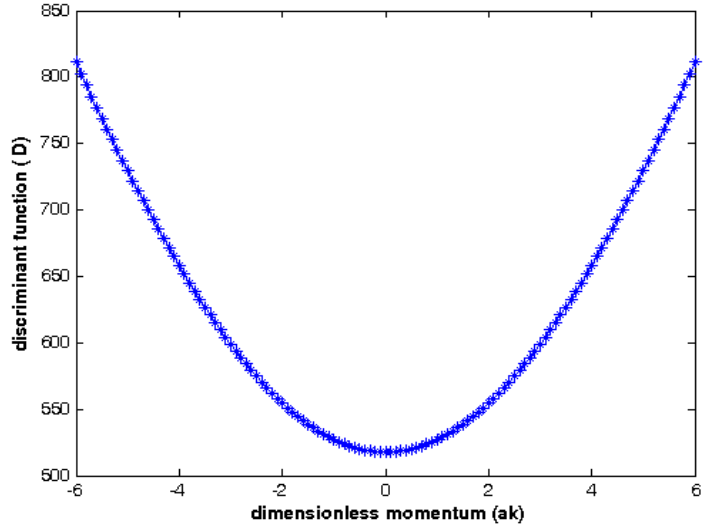

(a)

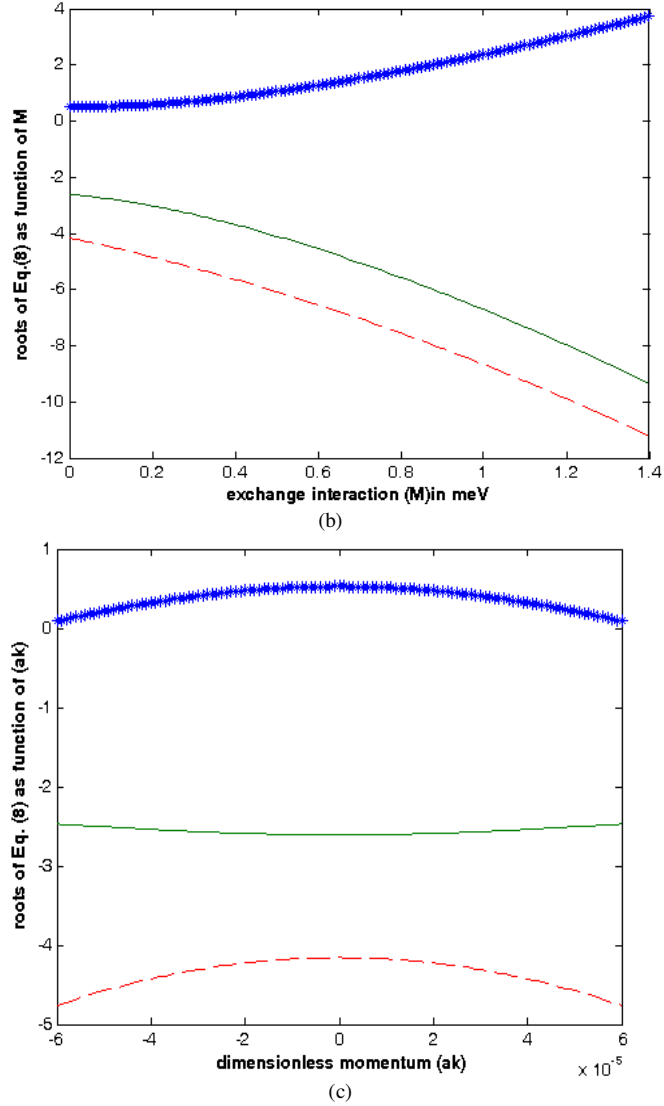

**Figure 2. (a)** A plot of the descriminant $\aleph$ as a function of $(a\delta k)$ at $M = 0$ for the Dirac point **K′**. **(b)** The plots of the three (real and distinct) roots of (8) as functions of $M$ (in meV) at $(a\delta k) = 0$ for the point **K′**. The uppermost curve corresponds to the admissible root $z_0(a\delta k, M, \xi)$ as this is positive. **(c)** The plots of the three (real and distinct) roots of (8) as functions of $(a\delta k)$ for $M = 0$ (for the point **K′**). The blue line corresponds to $z_0(a\delta k, M, \xi)$. We find that $z_0(a\delta k, M, \xi) = z_0(-a\delta k, M, \xi)$. The plots correspond to graphene on $WSe_2$.

energies without the Zeeman-like term at $\delta k = 0$. We have plotted $\sigma\lambda_{-s}(\xi, M)$ as a function of $M$ (in meV) in Figure 3. The absolute magnitude $|\lambda_{-s}(\xi, M)|$ is found to be decreasing function of $M$ over a broad range of values (0−1meV). The band energies get spin-splitted beyond $M = 0.9$ meV. This is perhaps an indication of the system cross-over/evolution to Elliot-Yafet [37] spin relaxation, via the route of the greater occurances of the spin-flip scattering events, from a D'yakonov-Perel' [38] type of spin-relaxation complaint environment (M < 0.8 meV). It may be noted that whereas in the former the inversion symmetry is retained, in the latter this symmetry is broken. Moreover, since $\sigma\lambda_s(\xi, M)$ is positive(negative) for conduction (valence) band, $\lambda_{-s}(\xi, M)$ is positive for all values of

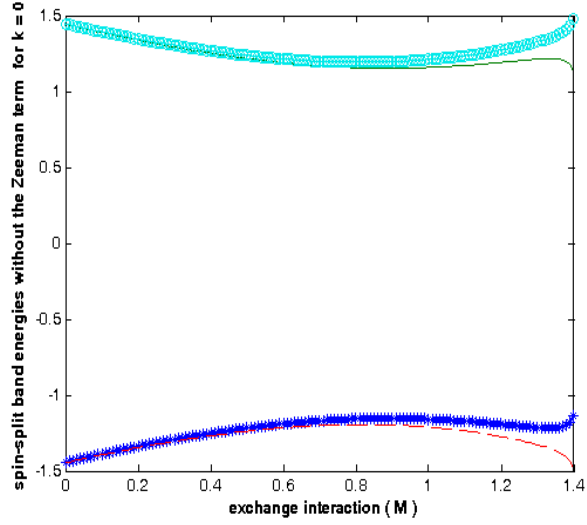

**Figure 3.** A plot of the spin-split valence and conduction band energies ($\sigma\lambda_s(\xi,M)$) without the Zeeman term ($s\sqrt{(z_0/2)}\ \lambda_R$) for wave vector k = 0 (at the Dirac point **K** ) as a function of the exchange interaction M in meV. The plot corresponds to graphene on WSe$_2$. For the Dirac point **K′** one obtains the similar plots.

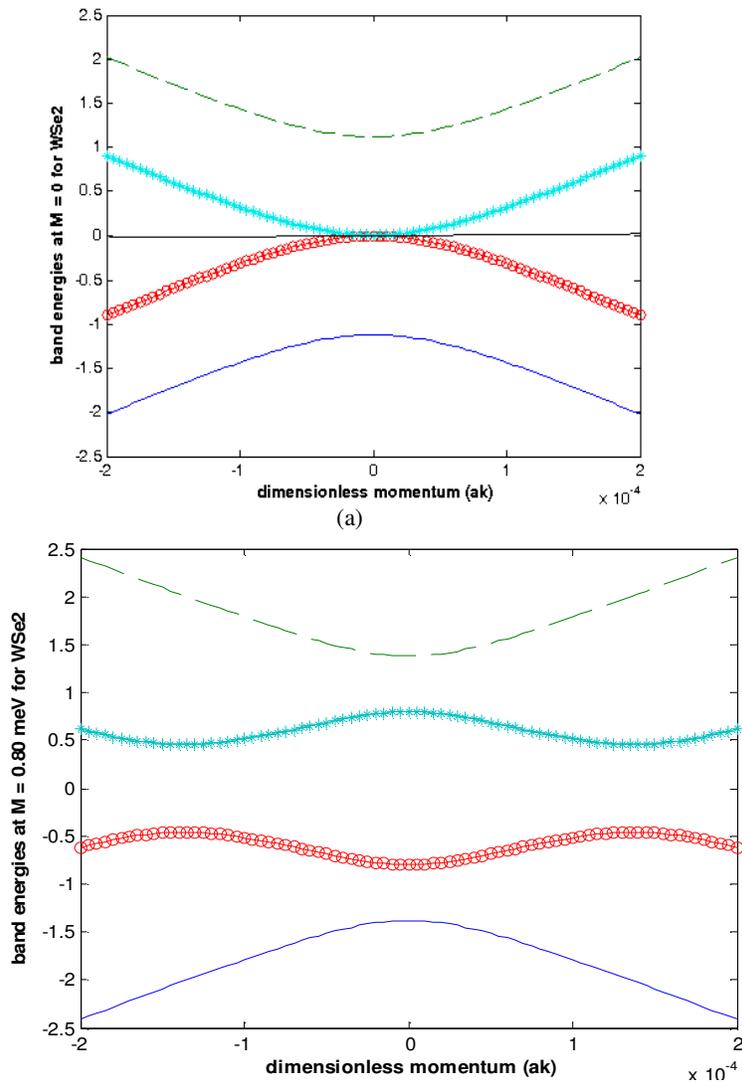

(a)

(b)

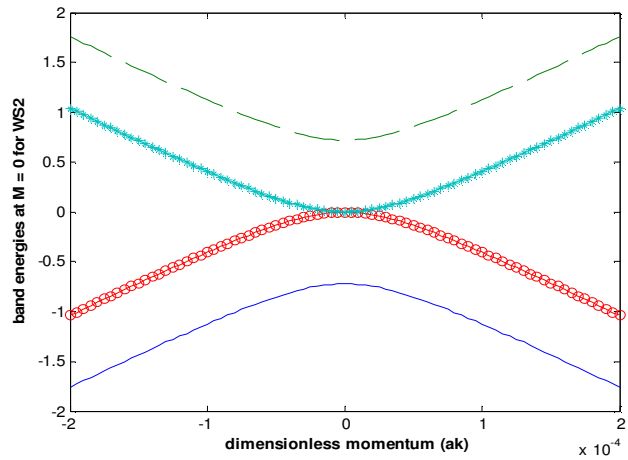

(c)

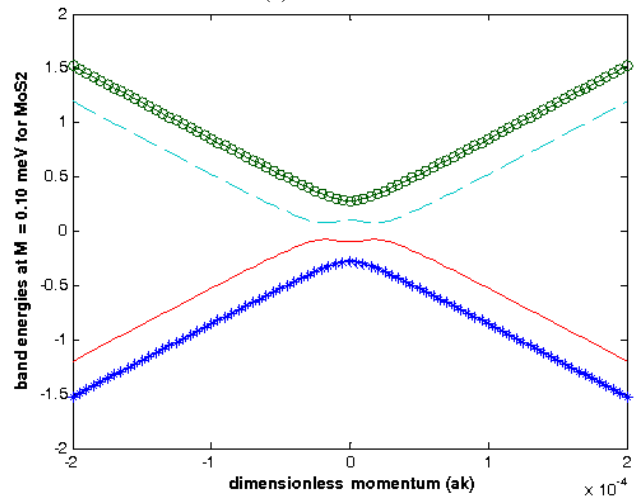

(d)

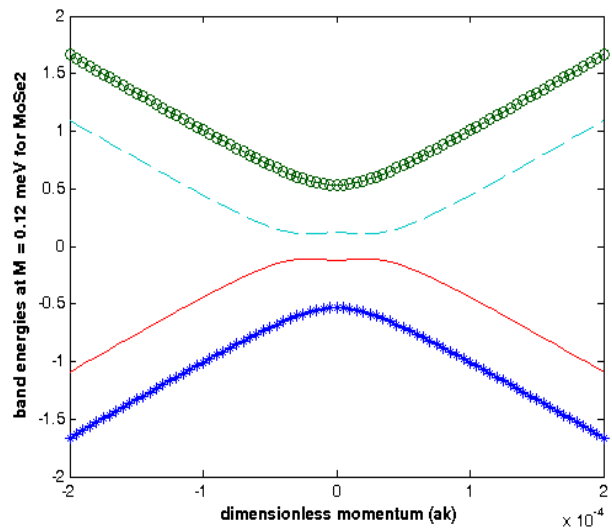

(e)

**Figure 4.** (a) A typical band structure at spin split semi-metallic phase corresponding to the special case of the dispersion in Eq.(10) near the **K** valley at M = 0 for graphene on WSe$_2$. (b) The band structure with a finite bulk gap corresponding to M = 0.80 meV. (c) A spin split semi-metallic phase bands for graphene on WS$_2$ at M = 0 around **K** point. (d) The anti-crossing of the non-parabolic bands with opposite spins around **K** point for graphene on MoS$_2$ at M = 0.10 meV. (e) The band structure with a finite bulk gap for MoSe$_2$ at M = 0.12 meV around **K** point. In all the plots the dispersion involves the RSOC and the exchange field (M) only.

*M*. We have shown typical band structure in Figure 4 corresponding to the spin split semi-metallic and the gapped phases of the simplified variant of the band-dispersion in Eq.(10) near the K valley for graphene on WSe$_2$, WS$_2$, MoS$_2$, and MoSe$_2$. The simplified variant comprises of the RSOC and the exchange field (M) only. The model is the same as that of the MacDonald et al. **[36]**. The figure (a) is for graphene on WSe$_2$ and M = 0. The figure (b) corresponds to M = 0.80 meV. Whereas figure (a) corresponds to a spin split semi-metallic phase, in figure (b) bulk band-gap develops. In figure (c), once again we have a semi-metallic phase (for graphene on WS$_2$ at M = 0) around **K** point. In figure (d), we have the anti-crossing of the non-parabolic bands in Eq.(10) with opposite spins around **K** point for graphene on MoS$_2$ at M = 0.10 meV. In figure (e), we have the band insulator regime where a finite bulk gap develops for MoSe$_2$ for M = 0.12 meV.

In Figure 5, we have shown the 2-D plots of the spin-split conduction and valence band energies in Eq.(10) for graphene on WSe$_2$ at the Dirac point **K** as a function of the dimensionless wave vector ($a|\delta k|$). We have considered the general case here, i.e taken into consideration the effect of the four substrate induced interactions together with the exchange field. We find that the graphene on TMDs is gapped at all possible exchange field values. On account of the strong spin-orbit coupling, the system acts as a quantum spin Hall insulator for $M = 0$. As the exchange field ($M$) increases, the band gap narrowing takes place followed by its recovery. The essential features of these curves, apart from the particle-hole symmetry, are (i) opening of an orbital gap due to the effective staggered potential, (ii) spin splitting of the bands due to the Rashba spin-orbit coupling and the exchange coupling, and (iii) the band gap narrowing and widening due to the many-body effect and the Moss-Burstein effect **[39]** respectively. The latter is due to the enhanced exchange effect. The exchange field *M* arises due to proximity coupling to ferromagnetic impurities, such as depositing Fe atoms to the graphene surface. Our plot in Figure 5 for the Dirac point **K** shows that as the exchange field increases the relevant band gap between the spin-down conduction band and the spin-up valence band gets narrower followed by the gap recovery and the gap widening. As regards MoY$_2$, we find that there is Moss-Burstein (MB) shift only and no band narrowing. Therefore, the exchange field could be used for the efficient tuning of the band gap in graphene on TMD. The shift due to the MB effect is usually observed due to the occupation of the higher energy levels in the conduction band from where the electron transition occurs instead of the conduction band minimum. On account of the MB effect, optical band gap is virtually shifted to high energies because of the high carrier density related band filling. This may occur with the elastic strain as well. Thus, studies are required to establish the simultaneous effect of the strain field and the carrier density on optical properties of GTMD. We note that the band gap narrowing and the $v_F$ renormalization, both, in Dirac systems, are essentially many body effects. Our observation of the gap narrowing in graphene on WSe$_2$, thus, supports the hypothesis of $v_F$ renormalization[40].Furthermore, (i) the direct information on the gap narrowing and the $v_F$ renormalization in graphene can be obtained from photoemission, which is a potent probe of many body effects in solids, and,(ii) as already mentioned, new mechanisms for achieving direct electric field control of ferromagnetism are highly desirable in the development of functional magnetic interfaces.

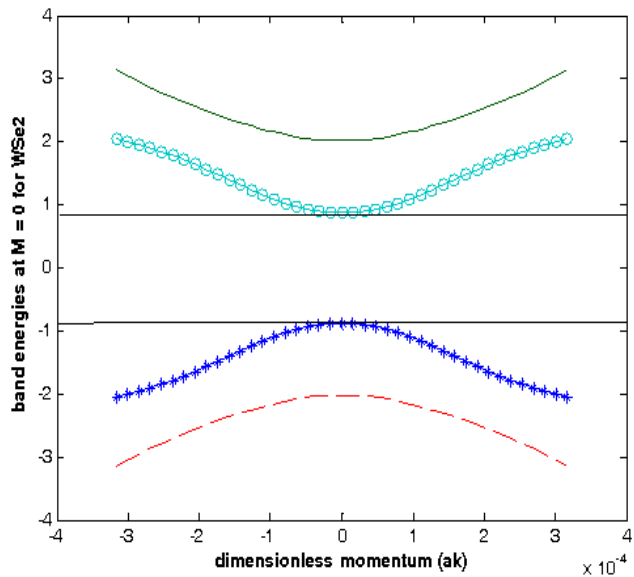

(a)

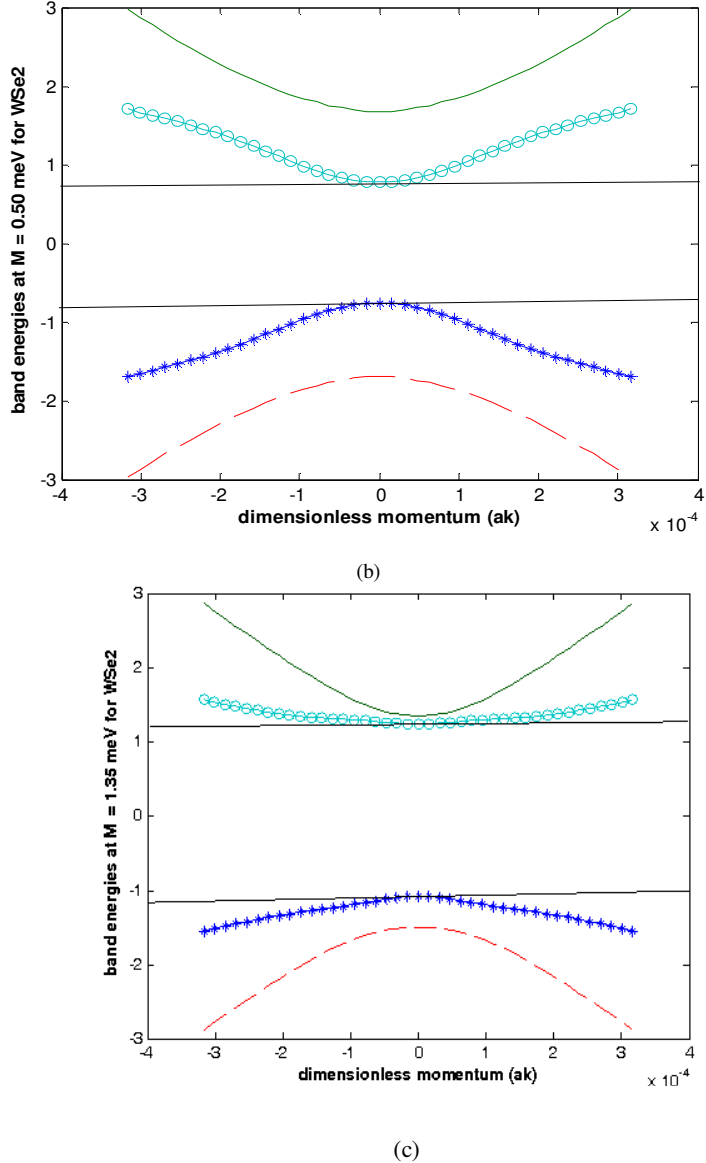

(b)

(c)

**Figure 5.** The 2-D plots of the spin-split conduction and valence band energies for graphene on WSe$_2$ at the Dirac point **K** as a function of the dimensionless wave vector $(a|\delta k|)$. The exchange field equal zero for (a). The field is 0.50 meV and 1.35 meV for (b) and (c), respectively.

### III. Polarization function

In this section, we examine the graphene plasmons. The plasmons are defined as longitudinal in-phase oscillation of all the carriers driven by the self-consistent electric field generated by the local variation in charge density. As an easily adaptable plasmonic material [**41,42**], graphene has attracted a lot of attention of the material science community. A tunable gate voltage applied to a graphene sheet is equivalent to the tuning of the chemical potential (μ) similar to that by the chemical doping. These processes provide a conduction band for the electrons which, in turn, supports plasmonic oscillations. The plasmonic properties of graphene can, thus, be controlled by tuning μ either by chemical doping or by the gating potentials.. Since the theoretical comprehension of plasmons in graphene on TMDC is far from complete, we examine here plasmon frequency in this system where the influence of the nunderlying substrate is mainly present in the single-particle excitation spectrum. We shall report in future intriguing effects generated in this system due to the many-body interaction , such as the excitations generated by electron–electron coupling (magnetoplasmons) and the composite modes arising from the coupling of plasmons with phonons and with charge carriers.

To find the full plasmon dispersion at finite wave vectors we need the quantum mechanical many-body theory for the collective motion of all carriers[43]. We start with the fact that the induced local charge density $\rho(r,\omega)$, giving rise to induced local potential $\Phi(r,\omega)$, corresponds to a dielectric effect. The many-body effect here changes the dielectric constant of the sample considerably even at the level of the one-loop contribution. We note that in a linear-response approximation, we have $\rho(\mathbf{r},\omega)=e^2\int d^2r'\chi(\mathbf{r},\mathbf{r}',\omega)\Phi(\mathbf{r}',\omega)$ where χ is the fermion response function or the dynamic polarization . This is a quantity of interest for many physical properties, since it determines e.g. the effective electron-electron interaction, the Friedel oscillations and the plasmon and phonon spectra. In terms of the bosonic Matsubara frequencies $\hbar\omega_n = 2\pi n/\beta$, the expression for the dynamic polarization function at finite temperature and finite chemical potential is given in refs. [43,44,45,46]. We now mention that the dimensionless Wigner-Seitz radius ($r_s$), which measures the ratio of the potential to the kinetic energy in an interacting quantum Coulomb system [47], is given in doped 2D graphene (a weakly interacting system) by a constant small compared to unity for all carrier densities. The strength of Coulomb interactions in graphene is determined by this dimensionless parameter. Since the random phase approximation (RPA) [46,47] is asymptotically exact in the $r_s \ll 1$ limit (non-interacting limit), the approximation is reasonable for graphene. The past experiments[48] have, however, suggested that the RPA significantly underestimates the static dielectric function of graphene. In this approximation, one obtains the polarization function in the momentum space, in the long-wavelength limit, as

$$\chi(a\delta\mathbf{q},\omega') = \sum_{\xi,s,s',\sigma,\sigma'=\pm 1} \chi_{\xi,s,s',\sigma,\sigma'}(a\delta\mathbf{q},\omega')$$

$$\chi(a\delta\mathbf{q},\omega') = g_v \sum_{\delta k, s,s', \sigma,\sigma' = \pm 1} |\langle \Psi_{s,\sigma}(a(\delta\mathbf{k}-\delta\mathbf{q}),M) | \Psi_{s',\sigma'}(a\delta\mathbf{k},M)\rangle|^2$$

$$\times [\frac{n_{s,\sigma}(a\delta\mathbf{k}-a\delta\mathbf{q},M)-n_{s',\sigma'}(a\delta\mathbf{k},M)}{\{\hbar\omega'+\varepsilon_{\xi=1,s,\sigma}(a(\delta\mathbf{k}-\delta\mathbf{q}),M) - \varepsilon_{\xi=1,s',\sigma'}(a\delta\mathbf{k},M) +i\eta\}}], \quad (11)$$

where $(\hbar\omega/\hbar v_F a^{-1})= \hbar\omega'$. Since we focus on the long wavelength plasmons here, we neglect transitions between two Dirac nodes located at different momenta. Within this assumption, contributions from other Dirac node in (12) can be taken into account replacing $\sum_\xi$ by the degeneracy factor $g_v$ and putting the valley index $\xi = +1$ in the summand. Here, $\varepsilon$ and $\psi$ are single-particle energies and wave funtions, and $n_{\xi,s,\sigma}(a\delta\mathbf{k},M) = [exp(\beta(\varepsilon_{\xi,s,\sigma}(a\delta\mathbf{k},M)-\mu')) +1]^{-1}$ is occupation function for the band σ = ± 1. The spin degenerate band-overlap of wave functions is given by $\mathcal{F}_{m,m'}(\delta\mathbf{k}\delta\mathbf{q}) =|\langle \Psi_{\xi,\sigma}(a\delta\mathbf{k}-a\delta\mathbf{q},M) | \Psi_{\xi,\sigma'}(a\delta\mathbf{k},M)\rangle|^2 = (1/2)(1+ mm'\cos\theta)$. The angle θ is that between states at $a(\delta\mathbf{k}-\delta\mathbf{q})$ and $a\delta\mathbf{k}$. Here, the indices $m$ and $m'$ denote the spin and all band quantum numbers for the occupied and empty states respectively. We note that the overlap of wave functions assume a simple form involving δ-function [43] for a non-chiral band structure. This excludes the possibility of any inter-band transitions between the non-chiral conduction and valence bands. For the graphene dispersion $E^\pm(\mathbf{k}) =\pm t \;|\phi_\mathbf{k}|-\mu$ with $\phi_\mathbf{k}=[1+2exp\;(i3ak_x/2)\;\cos(\sqrt{3}ak_y/2)\;]$, this overlap of wave functions, on the other hand, assumes the form

$$\mathcal{F}_{m,m'}(\delta\mathbf{k}\delta\mathbf{q}) = \tfrac{1}{2}[1+ mm'Re(exp(iaq_x)\frac{\phi_{\mathbf{k}-\mathbf{q}}\phi^*_\mathbf{k}}{|\phi_{\mathbf{k}-\mathbf{q}}||\phi_\mathbf{k}|})]. \quad (12)$$

We assume near zero temperature scenario for the moment. The occupation function $n_{\xi,s,\sigma}$ for T ≠0 K turns into a simple step function $\theta(\mu'-\varepsilon_{\xi,s,\sigma}(a\delta\mathbf{k},M\;))$ if T = 0 K, where $\mu'=\mu/(\hbar v_F/a)$ is the dimensionless chemical potential of the fermion number. All states below μ are occupied. If the sum over states k is understood as an integral over all one-particle states, the fermion density in d = 2 dimensions is $n \approx \frac{(k_F)^2}{\pi}$ , where $k_F$ is the Fermi momentum. In terms of the carrier concentration ($n$), the Fermi momentum in pure graphene is therefore $(ak_F) = a\sqrt{(\pi n)}$. The system, however, is with multi-band gapped energy spectrum given by $\varepsilon_{\xi,s,\sigma}(a|\delta k|,M) \approx [s\sqrt{(z_0(M)/2)\lambda_R} + \sigma\{(a|\delta k|)^2 + \lambda_{-s}(\xi,M)^2\}^{1/2}]$. The associated Fermi energy is $\varepsilon_F \approx \mu$ ($\varepsilon_{\xi,s,\sigma}(a|\delta k|,M) < \varepsilon_F$). This implies that the spin-dependent Fermi momentum is given by $|a\delta k| \leq (ak_{F,s}(\xi,M,\mu)) = \sqrt{\{(\mu - s\sqrt{(z_0/2)\lambda_R})^2 - \lambda_{-s}(\xi,M)^2\}}$. This is expression for $ak_{F,s}(\xi,M,\mu)$ in terms of the variable $\mu$. The Fermi momentum $ak_F = (\sum(ak_{F,s}(\xi,M,\mu))/2)$. The density of states (per unit area) of the system is given by $D(\varepsilon) \approx (2/\pi)(|\varepsilon_{\xi,s,\sigma}(a|\delta k|,M)|/(\hbar v_F)^2)$. At the Fermi energy, $D(\varepsilon_F) \sim \left(\frac{2}{\pi}\right)\sqrt{(\pi n)}/(\hbar v_F)$. In what follows, we write for the real part $\chi_1(a\delta q,\omega')$ of the integrand in (11) as

$$\frac{g_v(n_{\xi=1,m}(a\delta k - a\delta q,M) - n_{\xi=1,m'}(a\delta k,M))}{\{\hbar\omega' + \varepsilon_{\xi=1,m}(a(\delta k - \delta q),M) - \varepsilon_{\xi=1,m'}(a\delta k,M)\}} \mathcal{F}_{m,m'}(\delta k \delta q). \qquad (13)$$

The imaginary part is given by

$$\chi_2(a\delta q,\omega') = -g_v\pi\sum_{\delta k, m,m'=\pm 1}[n_m(a\delta k - a\delta q,M) - n_{m'}(a\delta k,M)] \delta(\hbar\omega' + \varepsilon_{\xi=1,m}(a(\delta k - \delta q),M) - \varepsilon_{\xi=1,m'}(a\delta k,M)) \mathcal{F}_{m,m'}(\delta k \delta q). (13)$$

Here we have used the identity $lim_{\eta \to 0+}1/(x\pm i\eta) = P(1/x) \mp i\pi\delta(x)$ with $P$ as the principal part. In the long-wave length limit, the band structure in Eq.(10) yields

$$\left| \varepsilon_{\xi=1,m}(a(\delta k \pm \delta q),M) - \varepsilon_{\xi=1,m}(a\delta k,M) \right| \approx \left\{\frac{a^2(\delta q^2 \pm 2\delta q \cdot \delta k)}{2\lambda_{-m}(\xi=1,M)}\right\}. \quad (14)$$

Note that the index $'m'$ in $\lambda_{-m}(\xi = 1, M)$ actually stands for the spin quantum number ($s$). We shall use here the band structure given in Eq. (10) (which is made dimension-less dividing by $(\hbar v_F/a)$) to retain the features of the present gapped electron-hole system and find the plasmon dispersion. Furthermore, we do not consider the spin-flip transitions here as they require a spatially varying Rashba spin-orbit interaction or the addition of magnetic molecules. The former one, for example, are involved in the scattering events associated with non-magnetic, charged impurities, acting as the source of an electric field that extends through the graphene, which may be located in the substrate. More importantly, since we have taken into account magnetic impurities, we need to specify, upto what value of the effective exchange field created by the latter, our non-inclusion of the spin-flip scattering events is valid. We shall show below that this limit is $M \sim 1$meV. Within the random phase approximation (RPA), the plasmon dispersion is obtained by finding zeros of the dynamical dielectric(Lindhard)function, which is expressed as[43, 47-52]

$$e_{\xi,m}^{(1)}(a\delta q,\omega') = 1 - \frac{V(\delta q)}{\left(\frac{\hbar v_F}{a}\right)}\chi_{\xi,m}^{(1)}(a\delta q,\omega') \qquad (14a)$$

to the first-order in the electron-electron interaction. Here $V(\delta q) = (e^2/2\varepsilon_0\varepsilon_r\delta q)$ is the Fourier transform of the Coulomb potential in two dimensions, $V(r) = e^2/4\pi\varepsilon_0\varepsilon_r r$, and $\varepsilon_0$ the vacuum permittivity and $\varepsilon_r$ is the relative permittivity of the surrounding medium. This Lindhard function does not take into account interactions between electrons beyond RPA, impurities, and phonons. Since the seminal work of Bohm and Pines, and Lindhard [46,47], a significant amount of effort has gone into addressing two main problems of applying RPA to real electronic systems, viz. it fails to reduce to the well-known results of classical Boltzmann theory in the long-wavelength limit and it faces the $a\delta q = 2 ak_F$ non-analyticity problem in the opposite limit. We, however, work here in the long-wavelength limit and thus do not encounter the $a\delta q = 2 ak_F$ singularity problem.

**A. Thomas-Fermi (TF) wave number $a\kappa$**

In order to find first the screening wave number (in two dimensions) in RPA [46], one needs to calculate the dielectric function in RPA given above. We can split the polarization function into two parts corresponding to the intra-band transitions and the inter-band transitions, respectively: $\chi(a\delta q,\omega') = \chi_{intra}(a\delta q,\omega') + \chi_{inter}(a\delta q,\omega')$, where $\chi_{intra}(a\delta q,\omega') = [\chi_{++}(a\delta q,\omega') + \chi_{--}(a\delta q,\omega')]$ and $\chi_{inter}(a\delta q,\omega') = [\chi_{+-}(a\delta q,\omega') + \chi_{-+}(a\delta q,\omega')]$. Now with hole doping the Fermi surface shifts to a lower energy. As a result the inter-band transitions with transition energy below $2E_F$ become forbidden, and it leads to a decrease in higher frequency inter-band absorption. At the same time, the lower frequency free carrier absorption (i.e. intra-band transition) increases dramatically. For simplicity, in what follows we consider the intra-band transitions without spin-flip only and obtain an approximate expression of the Thomas-Fermi wave number. We consider the formal expression of the polarization function $\chi(a\delta q,\omega')$ which may be written as $\chi(a\,\delta\boldsymbol{q},\omega') = \chi_1(a\,\delta\boldsymbol{q},\omega') + i\chi_2(a\,\delta\boldsymbol{q},\omega')$ where $\chi_1$ ($\chi_2$) is the real (imaginary) part of $\chi$. The static dielectric function, after a little algebra, may be written as

$$e(a\delta q, 0) = 1 - \frac{V(\delta q)}{\left(\frac{\hbar v_F}{a}\right)} g_v \sum_m \sum_{i,\delta k} (-a\delta q_i) \frac{\partial n_{\xi=1,m}(a\delta \boldsymbol{k},M)}{\partial(a\delta k)} \times \{\varepsilon_{\xi=1,m}(a(\delta\boldsymbol{k}-\delta\boldsymbol{q}),M) - \varepsilon_{\xi=1,m}(a\delta\boldsymbol{k},M)\}^{-1} \mathcal{F}_{mm}(\delta\boldsymbol{k},\delta\boldsymbol{q}),$$

where $e(a\delta q, \omega') = e_1(a\delta q, \omega') + ie_2(a\delta q, \omega')$. In the case of the electrostatic or chemical doping, we have

$$\sum_i (-a\delta q_i) \frac{\partial n_{\xi=1,m}(a\delta\boldsymbol{k},M)}{\partial(a\delta k)} = \sum_i (a\delta q_i) \frac{\partial n_{\xi=1,m}(a\delta\boldsymbol{k},M)}{\partial \mu'} \frac{\partial \varepsilon_{\xi=1,m}}{\partial(a\delta k)}$$

where $\frac{\partial \varepsilon_{\xi=1,m}(a\delta\boldsymbol{k},M)}{\partial(\delta k)} = \sigma \frac{(a|\delta\boldsymbol{k}|)}{\{(a|\delta\boldsymbol{k}|)^2 + \lambda_{-m}(\xi=1,M)^2\}^{1/2}}$. While the index '$m$' in $\lambda_{-m}(\xi=1,M)$ and $\varepsilon_{\xi=1,m}(a\delta\boldsymbol{k},M)$ stands for the spin quantum number ($s$) as already mentioned, the index '$\sigma$' stands for the band quantum number. We can thus write for the doped case

$$e(a\delta q, 0) \approx 1 + g_v \sum_m \left(\frac{\lambda_{-m} V(\delta q)}{\left(\frac{\hbar v_F}{a}\right)}\right) \frac{\partial}{\partial \mu'} \sum_{i,\delta k} (a\delta q_i) \frac{(a|\delta\boldsymbol{k}|) n_{\xi=1,m}(a\delta\boldsymbol{k},M)}{\{(a|\delta\boldsymbol{k}|)^2 + \lambda_{-m}(\xi=1,M)^2\}^{1/2}} \times (a\delta\boldsymbol{q}\cdot a\delta\boldsymbol{k})^{-1} \mathcal{F}_{m,m}(\delta\boldsymbol{k},\delta\boldsymbol{q}).$$

In the long wavelength limit $\lambda_{-m} \frac{(a|\delta\boldsymbol{k}|)}{\{(a|\delta\boldsymbol{k}|)^2 + \lambda_{-m}(\xi=1,M)^2\}^{1/2}} \approx (a|\delta\boldsymbol{k}|)$. Furthermore, $(a\delta\boldsymbol{q}\cdot a\delta\boldsymbol{k})$ is equal to $\sum_j (a\delta q_j)(a\delta k_j)$. One may now write approximately the statically screened coulomb potential as $V_s(\delta q, \omega'=0) = \frac{V(\delta q)}{e(\delta q, \omega'=0)}$ where $e(\delta q, \omega'=0) = 1 + (\kappa/\delta q)$. The Thomas-Fermi wave number ($\kappa$) is given by

$$\kappa \sim e^2 q_F / (2\hbar v_F \varepsilon_0 \varepsilon_r) = 2\pi \alpha\, c\, q_F / \varepsilon_r v_F.$$

Here $\alpha = \frac{e^2}{4\pi\varepsilon_0 \hbar c} = \frac{1}{137}$ is the fine-structure constant. For the comparison sake, one may write approximately the statically screened coulomb potential for the undoped graphene ($n=0$, $E_F=0$), which behaves as a zero-gap semiconductor, as $V_{s0}(\delta q, \omega'=0) = \frac{V(\delta q)}{e_0(\delta q, \omega'=0)}$ where $e_0(\delta q, 0) = e_0(0,0) = 1 + \frac{\pi}{2}\alpha_1$, $\alpha_1 = \frac{e^2}{4\pi \varepsilon_r \varepsilon_0 \hbar v_F} = \alpha\left(\frac{c}{\varepsilon_r v_F}\right)$ at zero-temperature. Therefore, it turns out that $e(\delta q, \omega'=0) = 1 + 2(e_0(0,0) - 1)\left(\frac{q_F}{\delta q}\right)$ and $V_s(\delta q, \omega'=0) \approx (2\pi\alpha \hbar c/\varepsilon_r)/(\delta q + 2.27 q_F)$. The quantity ($q_F$) is given by the derivative

$$g_v \sum_{\delta k} \sum_m \frac{\partial\left(n_{\xi=1,m}(a\delta\boldsymbol{k},M)\right)}{\partial \mu'} \mathcal{F}_{m,m}(\delta\boldsymbol{k},\delta\boldsymbol{q}) \qquad (15)$$

where the sum $g_v \sum_m \sum_{\delta k} n_{\xi=1,m}(a\delta\boldsymbol{k},M)$ is the charge density ($n$) multiplied by the sheet-area, and $n \approx \frac{(k_F)^2}{\pi}$. Note that, in the zero-temperature limit, the derivative $\frac{\partial n_{\xi=1,m}}{\partial \mu'}$ corresponds to $\delta\left(\mu' - \varepsilon_{\xi=1,m}(a\delta\boldsymbol{k},M)\right)$. One, however, finds straightway $q_F = k_F$ to the leading order in $\delta q$. To explain, we first note that for the graphene dispersion, the band-overlap of wave functions in (12) is a number $\sim 1$ to the leading order in the long wavelength limit. It suffices to use this quantity for general $\mathcal{F}_{m,m'}(\boldsymbol{k},\boldsymbol{q})$ here as the substrate induced interactions considered are in the nature of perturbations. Next, we inter-change the derivative $\frac{\partial}{\partial \mu'}$ with summations in (15) and approximate $g_v \sum_{\delta k,m} n_{\xi=1,m}(a\delta\boldsymbol{k},M)\mathcal{F}_{m,m}(\delta\boldsymbol{k},\delta\boldsymbol{q})$ by the carrier density $n$ to the leading order in $\delta q$. The term $q_F$ will then correspond to $\left(\frac{\partial n}{\partial \mu'}\right)$ to the leading order in $\delta q$. Since

$$k_F = \left(\frac{1}{2}\right)\sum k_{F,m}(\xi,M,\mu) = \left(\frac{1}{2}\right)\sum \sqrt{\{(\mu' - m\sqrt{\left(\frac{z_0(M)}{2}\right)}\lambda_R)^2 - \lambda_{-m}(\xi,M)^2\}},$$

we have $\left(\frac{\partial n}{\partial \mu'}\right) \approx \frac{\partial}{\partial \mu'} \frac{(\sum(k_{F,m}))^2}{4\pi}$. We find that

$$q_F \approx \left(\frac{k_F}{\pi}\right)\sum \left\{\frac{1}{\sqrt{[1-\{\frac{\lambda_{-m}(\xi M)^2}{\left(\mu'-m\sqrt{\left(\frac{z_0(M)}{2}\right)}\lambda_R\right)^2}\}]}}\right\} \; ; \; \kappa/k_F \sim \left(\frac{2\alpha c}{v_F \varepsilon_r}\right)\sum \left\{\frac{1}{\sqrt{[1-\{\frac{\lambda_{-m}(\xi M)^2}{\left(\mu'-m\sqrt{\left(\frac{z_0(M)}{2}\right)}\lambda_R\right)^2}\}]}}\right\} \quad (16)$$

The quantity $\kappa/k_F$ corresponds to the relative strength of screening. The immediate implication is that ($a\kappa$) and $\kappa/k_F$ could be changed by the tuning of the chemical potential and the exchange field. We also find that our analysis is valid as long as $|\mu' - m\sqrt{\left(\frac{z_0(M)}{2}\right)}\lambda_R| > |\lambda_{-m}(\xi,M)|$. That is, the chemical potential is greater than the effective gap parameter. The Fermi level lies outside the gap. The situation is basically corresponding to the long wave-length $((a|\delta\mathbf{q}|) \ll 1)$ limit. In this limit the energy and momentum are small, and therefore the limit is compatible with the low energy description of the system given above. In view of the values of $|\lambda_{-m}(\xi,M)|$ in Figure 3, this condition implies that the Thomas-Fermi screening and the plasmon frequency solution given below in the long wave-length limit is possible only when $|\mu|$ is greater than a certain $\mu_c(M)$ in the doped case. It may be noted that in the undoped case low-energy plasmon resonances are not possible. Only heavily damped high energy resonances($\pi$ plasmons) associated transitions between van Hove singularities are possible. We notice from Eq.(16) that stronger RSOC leads to larger $q_F$ and $\kappa$. The stronger RSOC, therefore, has foiling effect on the TF screening length. This could be seen from Figure 6. The RSOC parameter can be tuned by a transverse electric field and vertical strain, as already mentioned in section II. Interestingly, from the plots we also notice that the sreening length is greater when the exchange field strength is greater. The plots refer to the graphene on $WSe_2$. The chemical potential is assumed to be constant and equal to 6.6743 meV. As regards the ratio $\eta$ of the effective gap parameter to a constant electro-chemical potential, a higher value of $\eta$ yields higher value of $\kappa$. The foiling effect in Figure 6 (a), over a reasonably broad range of the exchange field values (0– 1meV), is possibly an indication of the domination of the spin meandering due to the effective magnetic field corresponding to the pseudo Zeeman term discussed in section II(Dyakonov-Perel (DP) mechanism [38,53]) in the system over the spin-flip scattering of electrons due to momentum scattering and spin-orbit splitting of the valence band (Elliot-Yafet (EY) spin relaxation mechanism [37,54]).The almost nil foiling effect for larger $M$ ($M$ >1meV) indicates establishment of the domination of the Elliot-Yafet mechanism due to increase in the spin relaxation rate via the introduction of more spin-flip scattering events. As the Rashba effect on screening length itself due to an external field is calculated to be rather weak[see also ref.**35**], the variation anticipated in Figure 6 may not be spectacular. In conventional 2D electron gases (2DEGs), the Thomas-Fermi wave vector $\kappa$ is generally independent of the carrier density. However, for the pure graphene the screening wave vector is proportional to the square root of the density. Thus, the relative strength of screening ($\kappa/k_F$), where $k_F \sim \sqrt{(\pi n)}$ is the Fermi wave-vector, in pure graphene is constant. In the large momentum transfer regime, of course, the static screening increases linearly with wave vector due to the inter-band transition [35]. For Gr-TMD, we obtain the similar result. As shown in Figure 6(b), the relative strength of screening is nearly a constant with relative to the changes in the gate voltage (or, the carrier density) and the exchange field strength; at lower densities the behavior is slightly contrary to what one would expect.

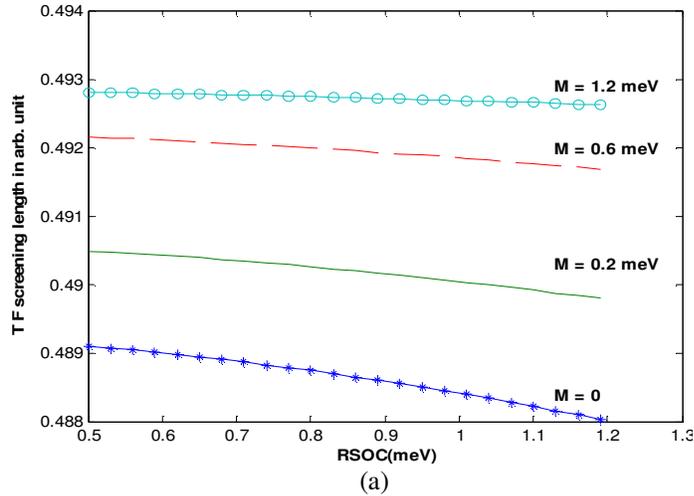

(a)

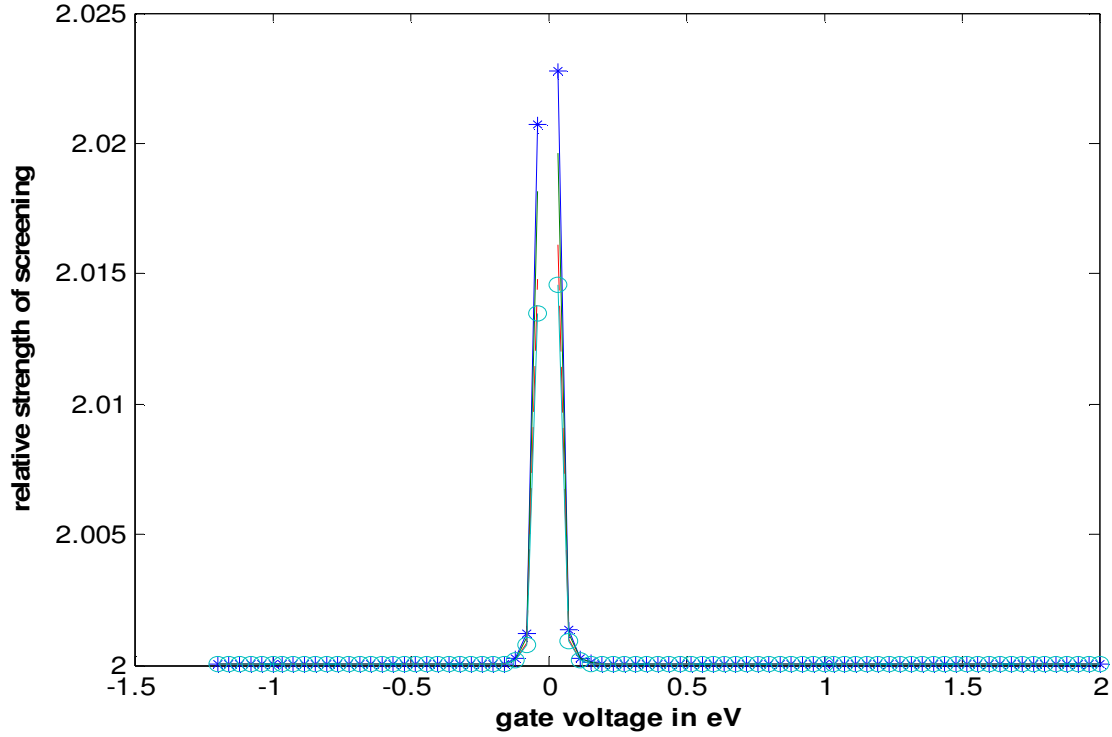

**Figure 6.** (a)The qualitative plots of the Thomas-Fermi (TF) screening length in arb. unit as a function of the RSOC parameter at a given value of the exchange field (M).The lowermost curve corresponds to M = 0 , and the two curves in the middle to M= 0.2meV and 0.6 meV . The uppermost curve corresponds to M= 1.2 meV. The plots refer to the graphene on $WSe_2$. The chemical potential is assumed to be constant and equal to 6.6743 meV.(b) The plots of the relative strength of the Thomas-Fermi (TF) screening as a function of the gate voltage.

**B. Plasmon frequency**

The plasmon dispersion can be obtained within the RPA by finding the zeros of this dielectric function (i.e. $e(\omega - i\gamma, a|\delta q|) = 0$ where $\gamma$ is the decay rate of plasmons). For weak damping, the equation $\text{Re } e(\omega, a|\delta q|) = 0$ yields the plasmon frequency $\hbar\omega_{pl}'$. As before, from Eqs. (13) and (14) we have

$$e_1(a\delta q, \omega') \approx 1 - g_v \sum_{m,\sigma} \left(\frac{\frac{e^2}{2\varepsilon_0 \varepsilon_r \delta q}}{\left(\frac{\hbar v_F}{a}\right)}\right)(\hbar\omega')^{-1} \sum_{i,\delta k}(a\delta q_i) \frac{\partial n_{\xi,=1,m}(a\delta k, M)}{\partial \mu'} \mathcal{F}_{m,m}(\delta k, \delta q)$$
$$\times \left(\sigma \frac{(a|\delta k|)}{\{(a|\delta k|)^2 + \lambda_{-m}(\xi=1,M)^2\}^{1/2}}\right) \times \left(1 - \sigma \left\{\frac{(a\delta q. a\delta k)}{(\hbar\omega')\lambda_{-m}(\xi=1,M)}\right\}\right)^{-1}. \quad (17)$$

Whereas the index $'m'$ in $\lambda_{-m}(\xi,=1, M)$ stands for the spin quantum number (s) as already mentioned, the index $'\sigma'$ stands for the band quantum number. Here the contributions from the two valley-states have been taken into account replacing $\sum_v$ by the degeneracy factor $g_v$ and putting the valley index $\xi = +1$ in the summand. We once again note that for the graphene dispersion, the band-overlap of wave functions in (12) is a number ~ 1 to the leading order in wave vectors and it suffices to use this over-lap for a general $\mathcal{F}_{m,m'}(\delta k, \delta q)$ here as the substrate induced interactions considered are in the nature of perturbations. However, in the quest for only the leading order terms in $\delta q$ in the plasmon dispersion, one may approximate the band-overlap by a number ~ 1. We now make use of the standard integral $\int_0^{2\pi} d\varphi/\{\vartheta - x \cos\varphi\} = 2\pi (2\theta(\vartheta)-1)(|\vartheta|^2 - x^2)^{-1/2}$, for $|\vartheta| > x$, and zero, for $|\vartheta| \le x$ in (17). Here $\theta(\vartheta)$ is the Heaviside step function. We can write

$$\int_0^{2\pi} d\varphi/\left[1 - \sigma\left\{\frac{a|\delta q|a|\delta k|)}{(\hbar\omega')\lambda_{-m}(\xi,M)}\right\}\cos(\varphi)\right] = 2\pi\left\{1 - \left(\frac{a|\delta q|a|\delta k|}{(\hbar\omega')\lambda_{-m}(\xi,M)}\right)^2\right\}^{-1/2}.$$

This integral allow us to write Eq.(17), for the conduction electrons, in the high frequency limit as

$$e_l(a\delta q,\omega') \approx 1-2\pi g_v \sum_i \int d(\delta k)\,(a\delta q_i)\sum_m \left(\frac{\frac{e^2 a^2}{2a\delta q\, a\varepsilon_0 \varepsilon_r}}{\left(\frac{\hbar v_F}{a}\right)}\right)(\hbar\omega')^{-1}\frac{\partial n_{\xi=1,m}(a\delta k,M)}{\partial \mu'}\left(\frac{(a|\delta k|)}{\{(a|\delta k|)^2 + \lambda_{-m}(\xi=1,M)^2\}^{1/2}}\right)$$
$$\times\left\{1-\left(\frac{a|\delta q|\,a|\delta k|}{(\hbar\omega')\,\lambda_{-m}(\xi,M)}\right)^2\right\}^{-1/2}$$

$$\approx 1 -\pi g_v \int d(\delta k)\sum_m \left(\frac{\frac{e^2 a^2}{2\,a\varepsilon_0\varepsilon_r}}{\left(\frac{\hbar v_F}{a}\right)}\right)(\hbar\omega')^{-3}\,\delta(\mu'-\varepsilon_{\xi=1,m}(a\delta k,M))\times\left(\frac{(a|\delta q|)^2 (a|\delta k|)^3}{\{(a|\delta k|)^2 +\lambda_{-m}(\xi=1,M)^2\}^{\frac{1}{2}}}\right)(\lambda_{-m}(\xi,M)^{-2}). \quad (18)$$

where $d(\delta k) = \left(\frac{d^2(\delta k)}{(2\pi)^2}\right)$ and $\frac{\partial n_{\xi=1,m}(a\delta k,M)}{\partial \mu'}$ is replaced by the delta function at T = 0K. In the long wavelength limit

$$(\lambda_{-m}(\xi,M)^{-2})\frac{1}{\{(a|\delta k|)^2 + \lambda_{-m}(\xi=1,M)^2\}^{\frac{1}{2}}} \approx \lambda_{-m}(\xi,M)^{-3}.$$

In this limit at T = 0 K, Eq. (18), therefore, could be written as

$$e_l(a\delta q,\omega') \approx 1 - C\left(\frac{(a|\delta q|)^{2/3}}{\hbar\omega'}\right)^3 \sum_m \frac{[(\mu'-m\sqrt{\left(\frac{z_0(M)}{2}\right)}\lambda_R)^2 - \lambda_{-m}(\xi,M)^2]^{\frac{3}{2}}}{\lambda_{-m}(\xi,M)^{-3}}. \quad (19)$$

where $C = \pi g_v\left(\frac{\frac{e^2}{2\,a\varepsilon_0\varepsilon_r}}{\left(\frac{\hbar v_F}{a}\right)}\right)$. The real part of the polarization function at zero temperature can now be written as

$$\chi_1(a|\delta q|,\omega') = \pi g_v \left(\frac{a|\delta q|}{\hbar\omega'}\right)^3 \sum_m \frac{[(\mu'-m\sqrt{\left(\frac{z_0(M)}{2}\right)}\lambda_R)^2 - \lambda_{-m}(\xi,M)^2]^{\frac{3}{2}}}{\lambda_{-m}(\xi,M)^{-3}}. \quad (20a)$$

The quantity $\lambda_{-m}(\xi=1,M)$ is the spin-split valence and conduction band energies dependent on $M$ (or the spectral gap function), without the Zeeman term $(m\sqrt{(z_0/2)}\,\lambda_R)$, at the Dirac point. We notice from (19) that the equation Re $e(\omega,a|\delta q|) = 0$ is a simple cubic in $\hbar\omega'$ with the only real solution given by

$$\hbar\omega'_{pl} = C^{1/3}\,(a|\delta q|)^{\frac{2}{3}}\{\sum_m \frac{[(\mu'-m\sqrt{\left(\frac{z_0(M)}{2}\right)}\lambda_R)^2 - \lambda_{-m}(\xi=1,M)^2]^{\frac{3}{2}}}{\lambda_{-m}(\xi,M)^{-3}}\}^{1/3}. \quad (20b)$$

At a finite temperature this solution and the real part of the polarization function may be written as

$$\hbar\omega'_{pl} = C^{1/3}(a|\delta q|)^{\frac{2}{3}}Q^{1/3}(\mu,T,M),\quad \chi_1(a|\delta q|,\omega') = \pi g_v\left(\frac{a|\delta q|}{\hbar\omega'}\right)^3 Q(\mu),$$

$$Q(\mu,T,M) = \sum_m \int d(a\delta k)\,\frac{\beta(a|\delta k|)^3}{4|(\lambda_{-m}(\xi=1,M))|^3}\cosh^{-2}\frac{\beta}{2}\left(\varepsilon_{\xi=1,m}(a\delta k,M)-\mu'\right). \quad (21)$$

Using a representation of the Dirac delta function, viz. $\delta(x) = \lim_{\grave{e}\to 0}(1/2\grave{e}\,\cosh^2\frac{x}{\grave{e}})$, Eq.(21) could be reduced to (20) in the zero-temperature limit. Therefore,

$$Q(\mu,T=0,M) = \{\sum_m \frac{[(\mu'-m\sqrt{\left(\frac{z_0(M)}{2}\right)}\lambda_R)^2 - \lambda_{-m}(\xi=1,M)^2]^{\frac{3}{2}}}{\lambda_{-m}(\xi,M)^{-3}}\}. \quad (22)$$

Upon including the full dispersion of graphene on TMDC ignoring the spin-flip mechanism completely, we thus find that there is only one collective mode and it corresponds to charge plasmons. As in the Thomas-Fermi screening case, the plasmon frequency solution given above in the long wave-length limit is possible only when $|\mu|$ is greater than a certain $\mu_c$. We find from Eq. (20) that the zero-temperature plasma frequency is increasing function of $\mu'$. Since $\left(ak_{F,s}(\xi,M,\mu)\right) = \sqrt{\{(\mu'-s\sqrt{\left(\frac{z_0(M)}{2}\right)}\lambda_R)^2 - \lambda_{-s}(\xi,M)^2\}}$, one may write the sum in (21) as

$$\sqrt[3]{\{\sum_s \left(\frac{ak_{F,s}(\xi=1,M,\mu)}{\lambda_{-s}(\xi=1,M)}\right)^3\}} \equiv \sqrt[3]{(a\frac{K_F}{\pi^2})} \sim n^{1/2}.$$

This yields the plasmon frequency for the present massive fermion at T = 0 K as

$$\omega_{m,pl}(|\delta\mathbf{q}|) = (v_F^2 e^2 g_v/2\hbar\varepsilon_0\varepsilon_r)^{1/3}(|\delta\mathbf{q}|)^{\frac{2}{3}} \sqrt[3]{(\frac{K_F}{\pi})}. \qquad (23)$$

Therefore, the power law dependence of the plasma frequency on wave vector is of the type $\omega_{m,pl}(|\delta\mathbf{q}|) \sim (a|\delta\mathbf{q}|)^{2/3}$; the carrier density dependence is approximately of the type $n^{1/2}$. We note that the fallout of the sensetivity to the TMD substrate is the non-validity of the well-known[49,50,55]dependences of the plasmon frequency, viz. $(a|\delta\mathbf{q}|)^{1/2}$ and $n^{1/4}$. Through extensive first-principles electronic structure calculations, it was shown[56] that both the out-of-plane ($\varepsilon_\perp$) and the in-plane ($\varepsilon_\parallel$) dielectric constants of graphene depend on the value of applied field. For example, $\varepsilon_\perp$ and $\varepsilon_\parallel$ are nearly constant (~3 and ~1.8, respectively) at low fields ($E_{ext} < 0.01$ V/Å) but increase at higher fields to values that are dependent on the system size. Upon taking the low-field value we find here $4\pi\varepsilon_0\varepsilon_r \sim 3 \times 10^{-10}$ N$^{-1}$-m$^{-2}$-Coul$^2$.With $v_F$= 0.5×10$^6$ m-s$^{-1}$, $|\delta\mathbf{q}| \sim k_F/50$, $k_F$~1.8×10$^8$ kg-m-s$^{-1}$ and areal carrier concentration $n$~1.0×10$^{16}$ m$^{-2}$, we then estimate the plasmon frequency as $\nu_{m,pl}(|\delta\mathbf{q}|)\sim$ 40 THz. The frequency belongs to the far-infrared region of the electromagnetic wave spectrum. It is gratifying to see that a finite chemical potential applied to a graphene sheet provides a conduction band for the electrons, allowing for plasmons supported by the graphene on TMDC. Unlike in standard semiconductors where the carrier type is fixed by chemical doping during the growth process, the Fermi level in graphene can be continuously driven between the valence and conduction bands simply by applying a gate voltage, i.e. electrostatic doping [57].We now invoke the relation(valid only when one is far away from the charge neutrality point (CNP)) given in ref. [57], between the external voltage($V_g$), the geometric gate-graphene capacitance per unit area, and the chemical potential($\mu$) for a gated graphene monolayer, given by

$$\mu/e = V_g - \Phi_0 - ep/C_g \qquad (24)$$

where $\Phi_0$ is the electrical potential attributed to the residual doping. The sign of the induced carriers is opposite to the sign of the applied gate voltage.Therefore, a negative (positive) gate voltage corresponds to the induced carrier as the hole (electron). At the value of p ~ 10$^{16}$ m$^{-2}$, the potential $\varnothing = (ep/C_g)$=13.33$eV$ for $C_g$ = 12nF/cm$^2$. This value of $C_g$ corresponds to gates based on oxide dielectrics. The capacitance of solid polymers based transparent gates are ~1 µF/ cm$^2$ which is two orders of magnitude larger than that of oxide dielectric gates. All these informations could be utilized in the expression for $\omega_{m,pl}$ in terms of $V_g$ and $p$:

$$\omega_{m,pl} = (v_F^2 e^2 g_v/2\hbar\varepsilon_0\varepsilon_r)^{1/3}(|\delta\mathbf{q}|)^{\frac{2}{3}}$$
$$\times \{\sum \frac{[\left(\frac{V_g-\Phi_0-\frac{(ep)}{C_g}}{\left(\frac{\hbar v_F}{a}\right)}-m\sqrt{\left(\frac{z_0(M)}{2}\right)}\lambda_R\right)^2 - \lambda_{-m}(\xi=1,M)^2]^{\frac{3}{2}}}{|(\lambda_{-m}(\xi,M)|^3}\}^{1/3} \qquad (25)$$

where the last line is valid for the cases far away from CNP. The plasmon frequency in the long wave-length limit, indeed, turns out to be gate-tunable for a given carrier density. For the graphical representation (shown in figure 7) purpose, we, however, use the well-known[58] general relation between $\mu$ and $V_g$ for a graphene-insulator-gate structure, viz.

$$\mu \approx \varepsilon_a [(m^2+2eV_g/\varepsilon_a)^{1/2}-m] \qquad (26)$$

where $m$ is the dimension-less ideality factor and $\varepsilon_a$ is the characteristic energy scale. The relation between $\mu$ and the carrier density may be given by $\mu \approx \hbar v_F \sqrt{(\pi|n|)} sgn(n)$ where $sgn(n) = \pm 1$ for electron(hole) doping. Our plots in Figure 7 show that the plasmon frequency increases (blue shift) with the increase in the absolute value of the gate voltage at a given value of the exchange field ($M$). However, with increase in $M$, the frequency shows a blue-shift followed by a red shift. We have taken the absolute value of the dimensionless wave vector to be 10$^{-4}$ and the dimension-less ideality factor as 5.0. In this figure, we have also plotted the (induced carrier density./(Fermi momentum)$^2$) as a function of the (gate-voltage/Fermi energy). A negative (positive) gate voltage corresponds to the induced carrier as the hole (electron).

One can now write the spectral function, $\aleph(\omega',\mu,T,\delta q) \equiv -V'(\delta q)^{-1}\text{Im}\{1/e(\omega,a|\delta q|)\}$, as

$$\aleph(\omega',\mu,T,\delta q) = V'(\delta q)^{-1}\, e_2(a\,\delta q,\omega')/(e_1^2(a\,\delta q,\omega') + e_2^2(a\,\delta q,\omega')), \tag{27}$$

where $V'(\delta q) = \mathcal{C}/\pi g_v(a\,\delta q)$. Close to the Plasmon frequency $\omega'_{m,pl}$, $e_1(a\,\delta q,\omega)$ and $e_2(a\,\delta q,\omega)$ may be written as

$$e_1(a\,\delta q,\omega') \approx -(\mathcal{C}/\pi g_v(a\,\delta q))(\omega' - \omega'_{m,pl})\, \partial_{\omega'}\chi_1(a\,\delta q,\omega')\big|_{\omega' = \omega'_{m,pl}},$$

$$e_2(a\,\delta q,\omega') = (\mathcal{C}/(a\,\delta q))\, K(a\delta q,\omega',T), \tag{28}$$

$$K(a\delta q,\omega',T) = \sum_{\delta k,m} [n_m(a\delta k - a\delta q, M) - n_m(a\delta k, M)]$$

$$\times\; \delta(\hbar\omega' + \varepsilon_{\xi=1,m}(a(\delta k - \delta q), M) - \varepsilon_{\xi=1,m}(a\delta k, M)). \tag{29}$$

From Eq.(21) we find that

$$\left(\partial_{\omega'}\chi_1(a\,\delta q,\omega')\big|_{\omega' = \omega'_{m,pl}}\right) = -(3\hbar\pi g_v/\mathcal{C}^{4/3})\left(\frac{a|\delta q|}{Q(\mu,T,M)}\right)^{\frac{1}{3}}. \tag{30}$$

Inserting Eqs. (28)-(30) into the expression for the spectral function we obtain

$$\aleph(\omega',\mu,T,\delta q) \approx \frac{(\pi g_v K(a\delta q,\omega',T))^{-1}}{\left[\left(\frac{\omega'_{m,pl} - \omega'}{\frac{\Gamma(\mu,\omega',T,\delta q)}{2}}\right)^2 + 1\right]}, \tag{31}$$

where $\Gamma(a\delta q,\mu,\omega',T) = (2/3)\,\mathcal{C}^{4/3}\,\left(\frac{(Q(\mu,T,M))^{\frac{1}{3}}}{\hbar(a|\delta q|)^{1/3}}\right) K(a\delta q,\omega',T)$. This resembles a Lorentzian with the full width at half maximum and height, respectively, given by the functions $\Gamma(\mu,\omega',T,\delta q)$ and $(\pi g_v K(a\delta q,\omega',T))^{-1}$. The spectral function $\aleph(\omega',\mu,T,\delta q)$ is symmetrical about the position of its maximum. This function characterizes the probability of electrons to undergo surface excitation in surface region. We note that the full width at half maximum could be controlled by changing the chemical potential through the electrostatic doping. In figure 7(a) we have shown the 2D plots of the plasmon frequency in arb.unit as a function of the gate voltage at different values of the exchange field. We have taken care not to assign larger values ($M > 0.1$meV) of the exchange field which may trigger spin-splitting ignored in our analysis. The plasmon frequency increases with the increase in the absolute value of the gate voltage at a given value of the exchange field. However, with $M$ the frequency shows a slight red-shift followed by a much larger blue-shift at a given gate voltage. The reason being the application of an electric field alters the carrier density so does the magnetic exchange interaction. The full width at half maximum as a function of the gate voltage at different values of the exchange field (not shown) qualitatively will be the same due to the presence of the term $Q(\mu,T=0,M)$. The 2D plots of the (induced carrier density./(Fermi momentum)$^2$) as a function of the (gate-voltage/Fermi energy) are shown in Figure 7(b). A negative (positive) gate voltage corresponds to the induced carrier as the hole (electron). The uppermost, the two curves in the middle, and the lowermost curves, respectively, correspond to the dimension-less ideality factor 5.0, 5.1, 5.2, and 5.3V. A contour plot showing the Plasmon frequency in arb. unit as a function of the gate voltage ($V_g$) and the absolute value of dimensionless wave vector in the case of graphene on WSe$_2$ at T ≈ 0 K is presented in Figure 7(c). The exchange field M = 0. The plot and the colour-bar indicate the increase in the Plasmon frequency with the increase in the absolute value of the gate votage at a given wave vector.

For comparison with the corresponding result for the pure graphene( i.e. without the substrate induced interactions), it is necessary to look at the small frequency limit ($\omega \ll v_F k_F$) in which case only long wavelength ($\delta q \ll k_F$) plasmons contribute to the scattering. In this case of the mass-less fermions, one can use a simple Drude model to

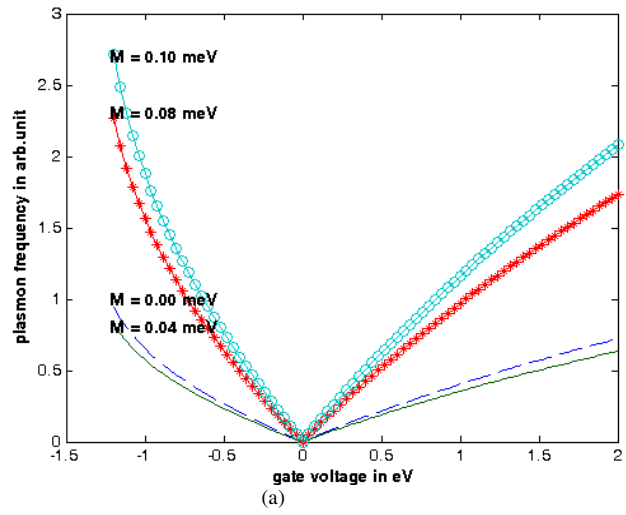

(a)

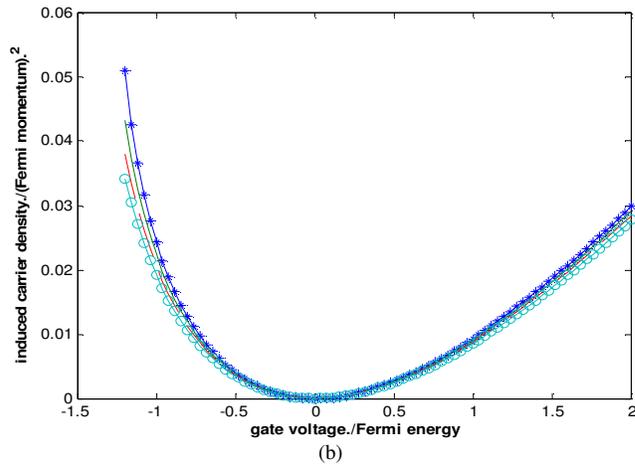

(b)

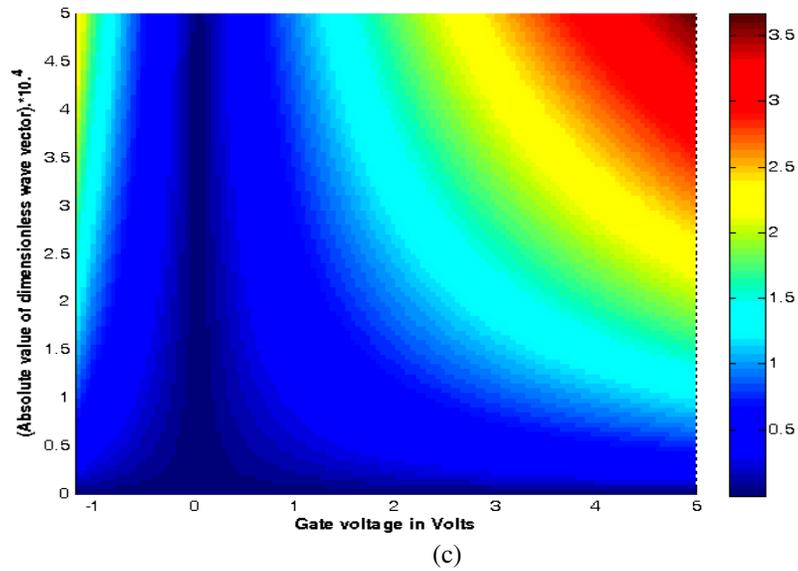

(c)

**Figure 7.** (a)The 2D plots of the plasmon frequency in arb.unit as a function of the gate voltage. The lowermost corresponds to the exchange field $M = 0.04$ meV, the two curves in the middle to M=0 and M = 0.08 meV, and uppermost to $M = 0.10$ meV. We have taken the absolute value of the dimensionless wave vector to be $10^{-4}$ and the dimension-less ideality factor as 5.0.The plasmon frequency decreases (red shift) with the increase in the absolute value of the gate voltage at a given value of the exchange field. However, with increase in $M$ the frequency shows a red-shift followed by a blue shift at a given gate voltage. (b) The 2D plots of the (induced carrier density./(Fermi momentum)$^2$) as a function of the (gate-voltage/Fermi energy).A negative (positive) gate voltage corres-ponds to the induced carrier as the hole (electron).The uppermost, the two curves in the middle, and the lowermost curves, respectively, correspond to the dimension- less ideality factor 5.0, 5.1, 5.2, and 5.3V. (c) A contour plot showing the Plasmon frequency in arb. unit as a function of the gate voltage ($V_g$) and the absolute value of dimensionless wave vector in the case of graphene on $WSe_2$ at T ≈ 0 K. The exchange field M = 0.The plot and the colour-bar indicate the increase in the Plasmon frequency with the increase in the absolute value of the gate votage at a given wave vector. Also, there is practically no increase in the Plasmon frequency with the increase in the absolute value of the dimensionless wave vector at a gate votage close to zero.

obtain the Plasmon frequency **[36]**:

$$\hbar\omega' = \left(\frac{\frac{e^2}{2a\varepsilon_0\varepsilon_r}}{\left(\frac{\hbar v_F}{a}\right)}\right)^{1/2} (a|\delta\boldsymbol{q}|)^{1/2} (ak_F/\pi)^{1/2}, \quad (32)$$

or,

$$\omega_{0,pl} = (v_F e^2/2\hbar\varepsilon_0\varepsilon_r)^{1/2} (|\delta\boldsymbol{q}|)^{\frac{1}{2}} \sqrt{\frac{k_F}{\pi}}. \quad (33)$$

It is evident that the power law dependence of the plasma frequency on wave vector is of the type $(\hbar\omega_{m,pl}') \sim (a|\delta\boldsymbol{q}|)^{2/3}$ for the gapped(massive) Dirac system, whereas for the ungapped graphene case it is $((\hbar\omega_{0,pl}') \sim (a|\delta\boldsymbol{q}|)^{1/2})$. Furthermore, whereas It may be mentioned **[29,33]** that, for the Drude–Sommerfeld model characterising the behaviour of electrons in a crystal structure of a metallic solid, the long wavelength plasmon dispersion is necessarily a classical plasma frequency with $(a|\delta\boldsymbol{q}|)^{1/2}$ dependence in 2D. In D dimensions, in general, the dependence is $((\hbar\omega_{0,pl}') \sim |\delta\boldsymbol{q}|^{(3-D)/2})$. Apart from this, there is a striking difference as well between the Dirac Plasma and the classical Plasma: Whereas, as one can see from Eqs.(21) and (23) that, the reduced Planck's constant appears explicitly in the leading term corresponding to the former, it appears in the sub-leading non-local corrections in the latter **[47,49]**. The factor of $1/\hbar^{1/3}$ ($1/\hbar^{1/2}$)explicitly appearing in the leading order term in the long wavelength plasmon dispersion of gapped (ungapped) graphene highlights their intrinsically quantum nature. In the case of stand-alone monolayer graphene, the corresponding results have been reported earlier **[49]**. One more important point is our complicated dependence of $(\hbar\omega_{m,pl}')$ on the carrier concentration $n$ ( approximately $n^{1/2}$) which is evident from the relations

$$\left(ak_{F,s}(\xi,M,\mu)\right) = \sqrt{(\mu' - s\sqrt{(z_0/2)}\lambda_R)^2 - \lambda_{-s}(\xi,M)^2} \quad (34)$$

and $(ak_F) \approx a\sqrt{(\pi n)}$. As regards the ungapped graphene, we find $\omega_{0,pl} \sim n^{1/4}$. This dependence is different from the standard two-dimensional electron gas's $n^{1/2}$ dependence. Thus, the substrate induced interactions change the fundamental character of the graphene plasmon by exhibiting $(|\delta\boldsymbol{q}|)^{\frac{2}{3}}$ and approximately $n^{1/2}$ dependences.

## IV. Discussion and concluding remarks

The possibility of the broadband mode confinement as well as associated field enhancement by the use of metal nanostructures involves the energy loss in metal via radiation absorption. This effect progressively increases with the field confinement **[59]**. A number of experimental **[60,61]** and theoretical**[62-66]** papers published so far have shown that plasmons in graphene corresponds to a deep sub-wavelength confinement resulting into a strong enhancement of the electromagnetic fields minus the enormous energy loss as in metal. We find that plasmon wavelength and graphene lattice constant($a$) ratio ($\lambda$) as a function of frequency ($f$) is given by $\lambda = K(n)f^{-3/2}$ where $K(n) \sim C\ n^{3/4}$. For $n \sim 10^{16}$ m$^{-3}$, and $a = 2.8 \times 10^{-10}$m, we find $a\ K(n) \sim 10^{12}$ m-Hz$^{3/2}$. This leads to the plasmon wavelength as 1μ-m at THz and $10^{-3}$μ-m at the mid infrared spectral range. Thus, in our case the plasmonic dispersion relation is of the form $f \sim const. \times n^{1/2}\ q^{2/3}$. In comparison, for the standalone graphene sheet, the plasmon wavelength is $J(n)f^{-2}$ where $J(n) \sim C_1\ n^{1/2}$. For the same value of the carrier density we find $J(n) \sim 10^{22}$ m-Hz$^2$. This leads to the plasmon wavelength as 10 mm at THz and 1μ-m at the mid infrared spectral range. The stronger confinement capability of Gr-TMD plasmon is obvious from above. The virtually loss-less, high field enhancement provided by graphene plasmons could be employed to increase the real part of the optical conductivity and the absorbance of this 2D material. The two quantities are closely related. In order to explain, we note down the formula for the real part of the impurity induced optical conductivity $Re\ \sigma(\omega)$ which could be derived with the aid of our results and that in ref. **[50]**, viz. $Re\ \sigma(\omega) = (n_i\sigma_0/2\pi)(q_{pl}/K_F\ q_F^2)(\omega_{m,pl}/\omega)^3$ where $q_{pl}= (\omega_{m,pl})^{3/2}(2\pi\hbar)^{1/2}(\varepsilon_0\varepsilon_r/v_F^2 e^2 K_F)^{1/2}$ is the wave vector at the Plasmon frequency, $\sigma_0=(e^2/4\hbar)$ the frequency-independent universal sheet conductivity of the massless Dirac fermions, and $n_i$ is the impurity concentration. Since $\omega_{m,pl}$ is an increasing function of the gate voltage(see Fig.7(a)), we find that the optical conductivity or the absorbance, too, is an increasing function. In Figure 8 we have plotted the absorbance in arb.unit as a function of the gate voltage for the various values of the exchange field. We find that the impurity induced absorbance is constant at lower values of the exchange field ($M < 0.10$ meV) with relative to the variation in the gate voltage. However, it increases as the exchange field ($M > 0.10$ meV) increases at a given gate voltage. We also find that, for a given value of the exchange field, the absorbance is constant at lower values of the the magnitude of the gate voltage. As the latter is assigned higher values, the absorbance increases.This outcome has potential for graphene-based optoelectronics. The universal minimal dc conductivity of the system being closely related to the optical conductivity, it is worthwhile to confirm the validity of the minimum value in the present case of the graphene on TMDC substrate.

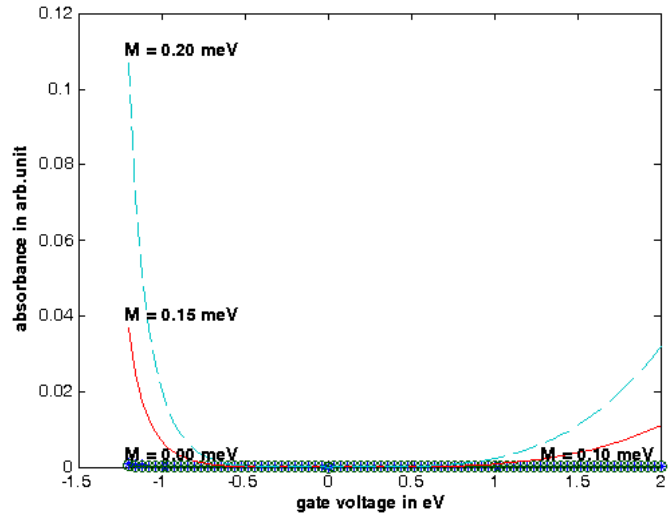

**Figure 8.** The 2D plots of the impurity induced absorbance in arb.unit as a function of the gate voltage at a given value of the exchange field (M). The lowermost curve corresponds to M = 0. Drawn in close proximity is another curve corresponding to M= 0.10 meV. While the curve in the middle corresponds to M= 0.15 meV, the uppermost curve corresponds to M= 0.25 meV. The plots refer to the graphene on $WSe_2$.

It may be mentioned that, recently, graphene plasmons in a graphene/insulator/metal heterostructure have been predicted to display a linear dispersion relation characterizing acoustic Plasmons and a considerably reduced wavelength, implying an improved field confinement. Alonso-González et al. **[67]**, reported real-space imaging of acoustic THz plasmons in a graphene photodetector with split-gate architecture. For this purpose they introduced nanoscale-resolved THz photo-current near-field microscopy, where near-field excited graphene plasmons are detected thermoelectrically rather than optically. The photocurrent images unveil strongly reduced plasmonic wavelengths, and a linear dispersion possibly resulting from the coupling of graphene plasmons with the metal gate below the graphene. The plasmon damping at positive carrier densities is dominated by Coulomb impurity scattering. In fact, a few years earlier the dispersion and the damping of the sheet plasmon in a graphene monolayer grown on Pt(111) had been studied by using angle-resolved electron energy loss spectroscopy. The investigators **[68,69]** discovered that the dispersion relation of the plasmon mode confined in the graphene sheet was linear which was conjectured to be a consequence of the screening by the metal substrate. This outcome demonstrated that the presence of an underlying metal substrate could have compelling ramifications on the plasmon propagation even in the case of a system which exhibits a weak graphene-substrate interaction. In our frame-work it is not difficult to demonstate that such plasmons are a real possibility when the carrier density (Fermi momentum) is large (areal carrier concentration $n\ 1.0\ 10^{17}$ m$^{-2}$ ). To this end, we recall the quantity of interest –the dynamical polarization –expressed in terms of the first order contribution in the electron–electron interaction in Eq.(14a). The self-consistent RPA result of the polarization to all orders in the electron–electron interaction is given by **[70]**

$$\chi_{\xi m}(a\delta\boldsymbol{q},\omega') = \chi_{\xi m}^{(1)}(a\delta\boldsymbol{q},\omega')/[1 - \frac{V(\delta q)}{\left(\frac{\hbar v_F}{a}\right)}\chi_{\xi m}^{(1)}(a\delta\boldsymbol{q},\omega')] \approx \chi_{\xi m}^{(1)}(a\delta\boldsymbol{q},\omega')/e(\delta q,\omega'=0)$$

$$= \chi_{\xi m}^{(1)}(a\delta\boldsymbol{q},\omega')/[1+(\kappa/\delta q)]$$

which leads to the plasmon dispersion obtainable, within the RPA, by finding the zeros of the self-consistent dielectric function

$$e_{\xi m}(a\delta\boldsymbol{q},\omega') = 1 - \frac{V(\delta q)}{\left(\frac{\hbar v_F}{a}\right)}\chi_{\xi m}(a\delta\boldsymbol{q},\omega') = 1 - \frac{V(\delta q)}{\left(\frac{\hbar v_F}{a}\right)}[\chi_{\xi m}^{(1)}(a\delta\boldsymbol{q},\omega')/\{1+(\kappa/\delta q)\}], \qquad (35)$$

where $V(\delta q) = (e^2/2\varepsilon_0\varepsilon_r\delta q)$ is the Fourier transform of the Coulomb potential in two dimensions. Upon substituting the expression for $\chi_{\xi m}^{(1)}(a\delta\boldsymbol{q},\omega')$ from Eq.(20a) in Eq.(35) we obtain the equation $e_{\xi m}(a\delta\boldsymbol{q},\omega') = 0$ in the form

$$0 = 1 - \frac{V(\delta q)}{\left(\frac{\hbar v_F}{a}\right)\{1+(\kappa/\delta q)\}}\pi g_v \left(\frac{a|\delta q|}{\hbar\omega'}\right)^3 \sum_m \frac{[(\mu' - m\sqrt{\left(\frac{z_0(M)}{2}\right)}\lambda_R)^2 - \lambda_{-m}(\xi M)^2]^{\frac{3}{2}}}{\lambda_{-m}(\xi M)^{-3}}. \qquad (36)$$

It is clear from (36) that, in the long wave length limit, for large areal carrier concentration ($n \geq 1.0 \times 10^{17}$ m$^{-2}$) $\{1 + (\kappa/\delta q)\} \approx (\kappa/\delta q)$ and therefore $\hbar\omega' \sim C(\mu') a|\delta\boldsymbol{q}|$ (acoustic plasmons). The tunability aspect of this type of plasmons is also obvious from (36) as $C = C(\mu')$.

The spin-valley dependent Zeeman term $[s\sqrt{(z_0(M,\xi)/2)}\lambda_R]$ of the dispersion in Eq.(10), obtained by us by calculaing the spectrum from a quartic, has its origin in the interplay of the substrate induced interactions with the prime player as the Rashba SOC. It is basically due the presence of the term $(4\varepsilon c)$ in Eq.(7) from the analytical view-point. The term mimics a real Zeeman field with opposite signs for the two physical spin states. Its non-triviality lies in the valley states and the exchange field dependence. This opens a gap at the neutrality point (see Figure 5). From arguments of band structure and $Z_2$ topological invariant, one may determine whether this gapped state is a topological insulator-a new quantum phase of matter that carries an odd number of helical edge states. A detailed investigation of the spin structure for the edge modes for the armchair geometry and the zigzag case is, however, needed to spell out the final verdict. Indeed, this spin-valley and RSOC dependent Zeeman term, together with the expectancy of electrically tuning RSOC and hence the band gap, unlocks the exciting possibility of further pursuit. It may be mentioned that Yang et al. **[71]** have recently predicted through their first-principle calculation that WS$_2$-covered graphene features a prominent 'valley-Zeeman' SOC that mimics a Zeeman field with opposite signs for the two valleys. Our analysis, however, did not yield this field. The reason could be traced to the fact that our model Hamiltonian does not involve a spin-orbit coupling($\lambda$) term of the type $(\lambda\tau_z s_z)$ where $\tau_z$ and $s_z$ act respectively on valley and spin degrees of freedom.

In the present communication we have introduced the magnetic exchange interaction in the most direct way using only the spin degrees of freedom. We have demonstrated that the exchange field can be used for efficient tuning of the band gap and the dielectric properties, such as the plasma frequency. The optical conductivity and absorbance, too, can be controlled by tuning of the exchange field as can be seen above. The field, in fact, plays a bigger role. It has been shown by Leutenantsmeyer et al. **[72]** that a spin current can be efficiently modulated at room temperature by controlling the exchange field. This is anticipated to usher in a new approach to regulate spins in the graphene based spintronic devices.

In conclusion, the Plasmons in graphene on TMD (GrTMD) have unusual properties and offer promising prospects for plasmonic applications covering a wide frequency range, ranging from terahertz up to the visible. Though immense progress has been made in graphene plasmonics over the past several years, never-the-less, we believe that our work casts new light onto the some of the important facts pertaining to the GTMD plasmonics. Looking backward, we observe there are many unsettled issues. For example, the problem of hybridization of plasmons with optical surface phonons that occurs when graphene is placed on a substrate has not been addressed. The graphene plasmons suffer from high losses at infrared wavelengths **[73]** attributed to the presence of multiple damping pathways **[74,75]**, such as collisions with impurities and phonons, as well as particle/hole generation via inter-band damping. We need to suggest the ways and means to curb these losses. Keeping in mind the rapid development in the field of graphene hybrid structures and alternative 2D materials and the general interest in investigating new possibilities thrown up by the plasmonics, spin-dependent physical properties, etc., doubtless, further revelations are waiting.